\documentclass[epj]{svjour}
\usepackage{epsfig} 
\usepackage{times}
\usepackage[active]{srcltx}

\begin{document}
\title{\hfill{\tiny FZJ-IKP-TH-2009-4,JLAB-THY-09-935}\\
\\
Neutral pion photoproduction at high energies}
%
\author{
A.~Sibirtsev\inst{1,2}, J.~Haidenbauer\inst{3,4}, S.~Krewald\inst{3,4},
U.-G.~Mei{\ss}ner\inst{1,3,4} and A.W. Thomas\inst{5,6}} 
\institute{
Helmholtz-Institut f\"ur Strahlen- und Kernphysik (Theorie) 
und Bethe Center for Theoretical Physics,
Universit\"at Bonn, D-53115 Bonn, Germany  \and
Excited Baryon Analysis Center (EBAC), Thomas Jefferson National
Accelerator
Facility, Newport News, Virginia 23606, USA
\and Institut f\"ur Kernphysik and J\"ulich Center for Hadron Physics,
Forschungszentrum J\"ulich, D-52425 J\"ulich, Germany
\and Institute of Advanced Simulations,
Forschungszentrum J\"ulich, D-52425 J\"ulich, Germany
\and Theory Center,  Thomas Jefferson National Accelerator
Facility, 12000 Jefferson Ave., Newport News, Virginia 23606, USA
\and College of William and Mary, Williamsburg, VA 23187, USA}
\date{Received: date / Revised version: date}

\abstract{
A Regge model with absorptive corrections is employed in 
a global analysis of the world data on the reactions
$\gamma{p}{\to}\pi^0p$ and $\gamma{n}{\to}\pi^0n$ 
for photon energies from 3 to 18~GeV. In this region
resonance contributions are expected to be negligible so that
the available experimental information on differential cross sections and
single- and double polarization observables at $-t{\leq}2$ GeV$^2$
allows us to determine the non-resonant part of the reaction 
amplitude reliably.
The model amplitude is then used to predict observables for photon 
energies below $3$ GeV. A detailed comparison with recent data from
the CLAS and CB-ELSA Collaborations in that energy
region is presented. 
Furthermore, the prospects for determining the $\pi^0$ radiative
decay width via the Primakoff effect from the reaction 
$\gamma{p}{\to}\pi^0p$ are explored.
}


\authorrunning{ A.~Sibirtsev {\it et. al} }
\titlerunning{Neutral pion photoproduction at high energies}

\maketitle
\section{Introduction}
Recently we have completed~\cite{Sibirtsev07} a systematic analysis of positive
and negative pion photoproduction at invariant collision energies $\sqrt{s}{>}$2
GeV. The main purpose of our study was to inspect whether the presently available data
indicate any evidence for excitations of baryons with masses up to 3~GeV 
or higher and to examine the issue of which observables might be most suitable 
for further pertinent investigations. 
The analyis was based on a Regge model with absorptive corrections that
took into account the $\rho$, $b_1$, and $a_2$ trajectories and pion exchange. 
Free parameters of the amplitudes were fixed in a global fit of the
world data on differential cross section and single polarization observables
available at high energies, i.e. at $\sqrt{s}{>}$3~GeV. In this region resonance
contributions are expected to be negligible so that the available experimental
information allows one to determine the non-resonant part of the reaction
amplitude reliably. The model amplitude was then used to predict observables at
energies below 3~GeV. Differences between the model predictions and data in this
energy region were systematically examined as possible signals for the presence
of excited baryons. 

In the present paper we extend this analysis to neutral pion photoproduction. 
Guided by our previous study of charged pion
photoproduction~\cite{Sibirtsev07} we use again a gauge invariant Regge model, 
which now comprises Regge pole and cut amplitudes for $\rho$, $\omega$,
and $b_1$ exchanges. Again, the free parameters of the model are fixed in a 
global fit to high energy data. Specifically, 
in our fit we include data on $\gamma{p}{\to}\pi^0{p}$ differential cross
sections and single and double polarization observables available in the energy
range 3${\leq}E_\gamma{\leq}$18 GeV but with the restriction $-t{\leq}2$ GeV$^2$.
Those data were obtained around or before 1980. 
After that we proceed to analyse data on neutral pion photoproduction on the 
proton collected recently by the CLAS Collaboration at JLab~\cite{Dugger} 
and by the CB-ELSA Collaboration in Bonn~\cite{Bartholomy,Pee} in the
energy region 2${\leq}E_\gamma{\leq}$3 GeV.

A strong motivation for studying neutral pion photoproduction is provided
by the occurence of the well-known Primakoff effect~\cite{Primakoff}. 
This effect is connected with the contribution of the one-photon exchange 
amplitude and manifests itself through a large differential cross section at 
very forward angles.
The one-photon exchange amplitude, in turn, is directly connected 
with the $\pi^0$ radiative decay width and proportional to the charge of
the target $Z$. The $\pi^0{\to}\gamma\gamma$ decay amplitude is related to 
symmetry breaking through the axial anomaly and reveals one of the 
fundamental properties of QCD 
\cite{Adler,Bell,Bardeen,Wess,Witten,Bijnens,Goity,Kampf}.

The fact that the differential cross section due to the Primakoff amplitude 
is proportional to $Z^2$ initiated strong activities in the determination 
of the $\pi^0$-meson radiative decay from measurements with nuclear
targets~\cite{Engelbrecht,Bellettini,Kryshkin1,Kryshkin2,Browman,Hadjimichael}.
Indeed,
most of the results~\cite{Bellettini,Kryshkin1,Kryshkin2,Browman} on the neutral
pion lifetime, given by the PDG~\cite{PDG}, were obtained by utilizing the 
Primakoff effect.
A very precise experiment (PrimEx) on the determination of the $\pi^0{\to}\gamma\gamma$
decay width from $\pi^0$-meson photoproduction on nuclear targets is
presently performed at JLab~\cite{Gasparian}. 

It was argued~\cite{Engelbrecht,Gasparian,Rodrigues1}, however, that pion
absorption on nuclei as well as the interference between the one-photon
exchange and the nuclear amplitude may complicate the data analysis and 
significantly affect the accuracy of the results obtained on the 
$\pi^0$-meson lifetime. 
While the one photon exchange amplitude is explicitly given by
theory, the evaluation of the nuclear amplitudes requires a precise 
knowledge of the elementary amplitude on a nucleon and the spectral 
function of the target nucleus.
Although 
the $Z^2$-argument clearly favors nuclear targets, this advantage might be
completely counterbalanced by the benefit of the $\pi^0$-meson photoproduction 
on the proton where one has a much better handle on the hadronic part of the
production amplitude. 
Therefore, in the present paper we will re-examine \cite{Laget} the 
prospects for determining the $\pi^0$ radiative decay width from neutral 
pion photoproduction on a proton target.  

The paper is organized as follows. In Sect.~2 we formulate the reaction
amplitudes. The parameters of the global fit to high energy data are given in
Sect.~3. The analysis of neutral pion photoproduction at high and low energies 
is given in Sects. 4 and 5, respectively. Results for the ratio of the 
$\gamma{n}{\to}\pi^0n$ and
$\gamma{p}{\to}\pi^0p$ differential cross section and the total cross section for
the neutral pion photoproduction from the proton are presented in Sect.~6.
In this section we also examine the energy dependence of the data for
fixed four-momentum transfer. 
In Sect.~7 we discuss the Primakoff effect on the proton target and 
explore the prospects for future experiments. 
The paper ends with a short Summary. 

\section{Reaction amplitudes}
Guided by our previous analysis of charged pion
photoproduction~\cite{Sibirtsev07} we use a gauge invariant Regge model which
combines the Regge pole and cut amplitudes for $\rho$, $\omega$ and $b_1$
exchanges.  At high energies the interactions before and after the basic
Regge pole exchange mechanisms are essentially elastic or diffractive scattering
described by Pomeron exchange. Such a scenario can be related to the distorted
wave approximation and provides a well defined
formulation~\cite{Irving,Irving1,Sopkovitch,Gottfried,Jackson3,Worden}
for constructing
Regge cut amplitudes. This approach, which can also be derived in an eikonal
formalism~\cite{Arnold} with $s$-channel unitarity~\cite{Jackson}, is followed 
in our work. Detailed discussions about the non-diffractive multiple scattering
corrections involving intermediate states which differ from the initial and
final states and the relevant Reggeon unitarity equations are given in
Refs.~\cite{Irving,Gribov,White3,White4}. For simplicity we do not consider
these much more involved mechanisms which would increase significantly the
number of parameters to be fitted.

We use the $t$-channel parity conserving helicity amplitudes $F_i$ ($i{=}
1,...,4$). Here $F_1$ and $F_2$ are the natural and unnatural spin-parity
$t$-channel amplitudes to all orders in $s$, respectively. $F_3$ and $F_4$ are
the natural and unnatural $t$-channel amplitudes to leading order in $s$. 

Each  Regge pole helicity amplitude is parameterized as
\begin{eqnarray}
F(s,t) =  \pi \beta(t)   \frac{1 {+} {\cal S}\exp[-i \pi \alpha
(t)]}{\sin[\pi\alpha(t)] \,\,\Gamma[\alpha(t)]}
\left[\frac{s}{s_0}\right]^{\alpha(t)-1},
\label{rpropa}
\end{eqnarray}
where $s$ is the invariant collision energy squared, $t$ is the squared 
four-momentum transfer and $s_0$=1~GeV$^2$ is a scaling parameter that
allows us to define a dimensionless amplitude. Furthermore, $\beta(t)$ is a 
residue function, 
${\cal S}$ is the signature factor and $\alpha(t)$ is the Regge trajectory. 

From Eq.~(\ref{rpropa}), we see that the factor $\sin[\pi\alpha(t)]$ would
generate poles at $t{\le}0$ when $\alpha(t)$ assumes the values $0,-1,\ldots$. 
The function $\Gamma[\alpha(t)]$ is introduced to suppress those 
poles that lie in the scattering region because 
\begin{eqnarray}
\Gamma[\alpha(t)] \Gamma[1-\alpha(t)]= \frac{\pi}{\sin [\pi \alpha(t)]}.
\label{eq:supp}
\end{eqnarray}

The structure of the vertex function $\beta(t)$ of Eq.~(\ref{rpropa})
is defined by the quantum numbers of the particles at the interaction
vertex, similar to the usual particle exchange Feyman diagram. 

Both natural and unnatural parity particles can be exchanged in the $t$-channel.
The naturalness $\cal N$ for natural (${\cal N}{=}{+1}$)  and unnatural (${\cal
N}{=}{-1}$)  parity exchanges is defined as
\begin{eqnarray}
{\cal N}=+1 \,\,\, \mathrm{if} \,\,\, P=(-1)^J, \nonumber \\
{\cal N}=-1 \,\,\, \mathrm{if} \,\,\, P=(-1)^{J+1},
\end{eqnarray}
where $P$ and $J$ are the parity and spin of the particle, respectively.
Furthermore, in Regge theory each exchange is denoted by a signature
factor ${\cal S}{=}{\pm}1$ defined as~\cite{Irving,Collins2,Collins3}
\begin{eqnarray}
{\cal S} = P \times {\cal N} = (-1)^J,
\label{signat}
\end{eqnarray}
which enters Eq.~(\ref{rpropa}).

To proceed further we should specify the trajectories that contribute to 
neutral pion photoproduction. Note that the mechanisms for charged and neutral
pion photoproduction are different. Indeed, for charged pion photoproduction
pion exchange dominates at small $-t$ while $\omega$-exchange is forbidden
altogether. For $\pi^0$-meson photoproduction just the opposite is the case. 

One of the significant differences between the data
on neutral and charged pion photoproduction is the presence of the dip in the
$\gamma{p}{\to}\pi^0p$ differential cross sections at the squared four-momentum
transfer $t{\simeq}$--0.5 GeV$^2$. A similar dip is also observed in other
reactions. For instance, in the $\pi^-p{\to}\pi^0n$ reaction such a dip results
from the $\rho$-exchange amplitude and its position is related to the
$\rho$-trajectory~\cite{Huang}. 

Therefore, one expects that $\rho$-exchange might dominate the
$\gamma{p}{\to}\pi^0p$ reaction too. Indeed, the square of the amplitude of
Eq.~(\ref{rpropa}) for the $\rho$-exchange is
proportional to
\begin{eqnarray}
|F(s,t)|^2 \propto 1 - \cos[\pi \alpha(t)],
\end{eqnarray}
which has a zero at $t{\simeq}$-0.6~GeV$^2$, when taking the $\rho$-trajectory
as obtained recently in a fit~\cite{Huang} to the available data for the 
$\pi^-p{\to}\pi^0n$ reaction. Note that the additional contributions to the
reaction amplitude ($\omega$, $b_1$) can move this zero closer to experimentally 
observed value.

\begin{table}[t]
\begin{center}
\caption{Correspondence between $t$-channel pole exchanges and the helicity
amplitudes $F_i$  ($i{=}1{\div}3$). Here $P$ is parity, $J$ the
spin, $I$ the isospin, $G$ the $G$-parity, $\cal N$ the naturalness and $\cal S$
the signature factor.}
\renewcommand{\arraystretch}{1.3}
\label{taba1}
\begin{tabular}{|c|c|c|c|c|c|c|c|}
\hline
$F_i$ & $P$  & $J$   & $I$  & $G$ & $\cal N$& $\cal S$& Exchange \\
\hline
$F_1$ & -1 & 1 & 1 & +1 & +1 & -1 &  $\rho$  \\
$F_1$ & -1 & 1 & 0 & +1 & +1 & -1 &  $\omega$  \\
$F_2$ & +1 &  1 & 1 & +1 & -1 & -1 & $b_1$ \\
$F_3$ & -1 & 1 & 1 & +1 & +1 & -1 &  $\rho$  \\
$F_3$ & -1 & 1 & 0 & +1 & +1 & -1 &  $\omega$  \\
\hline
\end{tabular}
\end{center}
\end{table}

In the $\gamma{p}{\to}\pi^0p$ reaction there is no difference between the
$\rho$ and $\omega$-exchanges if their trajectories are the same. Thus in
some previous studies~\cite{Barker1} both contributions were considered as 
just one exchange amplitude. However, in our study we treat $\rho$ and $\omega$
exchanges separately, because any possible difference in the amplitudes might
play a role in describing observables~\cite{Kellett,Guidal,Vanderhaeghen}. 
For instance it could allow us to fix the ratio of the $\gamma{n}{\to}\pi^0n$
and $\gamma{p}{\to}\pi^0p$ differential cross sections. Note that in these two
reactions the contributions from the isovector exchange enter with a different 
sign, whereas the isoscalar exchange is the same in both cases.

The contributions of the $\rho$ and $\omega$-exchanges to the reaction
amplitudes $F_i$ are indicated in Table~\ref{taba1} together with the relevant
quantum numbers.
Both $\rho$ and $\omega$ have natural parity and contribute to $F_1$ and $F_3$.
It was argued~\cite{Guidal} that if there are no other contributions one would
expect that the photon asymmetry $\Sigma$ would be predominantly +1. 
(However, 
note that $\Sigma$ vanishes at forward and backward directions~\cite{Nakayama}.) 
This can be easily understood when considering the relations
between the observables and the $t$-channel parity conserving helicity
amplitudes given in the Appendix~A. Experimental data~\cite{Bellenger,Anderson1}
available at high energies shows that the asymmetry indeed is consistent with
+1 for $t$ values above -0.4 and below -1.1. For -1${<}t{<}$-0.4 GeV$^2$ there
is a dip in the asymmetry and at $t{\simeq}$-0.5~GeV$^2$ it drops to
$\Sigma{\simeq}0.7$. Thus, one needs to take into account additional
trajectories that yield contributions to the unnatural-parity 
amplitudes $F_2$ and $F_4$.
Indeed, 
the leading $b_1$ trajectory, which we already used in the analysis of 
charged pion photoproduction contributes to $F_2$, cf. Table~\ref{taba1}, 
and, therefore, we include it here too. 

As we discussed in Ref.~\cite{Sibirtsev07} the empirical information about 
those trajectories that yield contributions to $F_4$ is very sparse. 
Hence, we neglected this amplitude in our analysis of charged 
pion photoproduction.
For the same reason we also neglect $F_4$ in the present study. Our
decision can be justified by considering the data~\cite{Booth,Bienlein,Deutsch} 
on the target ($T$) and recoil ($P$) asymmetry, available for the 
$\gamma{p}{\to}\pi^0p$ reaction at photon energies above 4 GeV. 
Since they fulfill roughly $T{\simeq}P$ one can conclude that $F_4{\simeq}$0, 
based on the relations given in Eqs.~(\ref{obs3}) and (\ref{obs4}) 
of Appendix~A. We will discuss this point below. 

The trajectories are taken in the following linear form,
\begin{eqnarray}
\alpha(t){=}\alpha_0+\alpha^\prime{t} \ , 
\label{eq:traj}
\end{eqnarray}
where the intercept and slope for the  $\rho$ and $b_1$ trajectories are taken 
over from analyses of other
reactions~\cite{Sibirtsev07,Irving,Huang,Sibirtsev2}.
Explicitly we have for the  $\rho$ and $b_1$ trajectories 
\begin{eqnarray}
\alpha_{\rho} &=& 0.53 + 0.8 \, t \ , \nonumber \\
\alpha_{b_1} &=&  0.51 + 0.8\, t \ .
\label{traj1}
\end{eqnarray}
The $\omega$-trajectory was parameterized by Eq.~(\ref{eq:traj}) with the slope
$\alpha^\prime$=0.8 GeV$^{-2}$, {\it i. e.} the same value as for other 
trajectories. The parameter $\alpha_0$ for $\omega$-exchange was fixed by 
a fit to the data.

The residue functions $\beta(t)$  used in our analysis are compiled in
Table~\ref{tab0}. They are similar to the ones used in some of the previous
analyses~\cite{Kellett,Rahnama2}. These residues are slightly different from
those applied in our study of charged pion photoproduction~\cite{Sibirtsev07}
because here we use explicitely the $\Gamma$ function in the amplitude
parameterization of Eq.~(\ref{rpropa}).

\begin{table}[t]
\begin{center}
\caption{Parameterization of the  residue functions $\beta(t)$ for the
amplitudes $F_i$, \, ($i{=}1,2,3$). Here $c_ {ij}$ is the coupling constant 
where the double index refers to the amplitude $i$ and the type of exchange
$j$, as specified in the Table.}
\renewcommand{\arraystretch}{1.3}
\label{tab0}
\begin{tabular}{|l|l|c|l|}
\hline
     & $\beta(t)$ & Exchange & $j$ \\
\hline
\multicolumn{4}{|c|}{ Pole amplitudes }\\
\hline
$F_1$ & $c_{11}$  & $\rho$ & 1 \\ 
$F_1$ & $c_{12}$  & $\omega$ & 2 \\ 
\hline
$F_2$ & $c_{23}\, t $  &
$b_1$ & 3 \\ 
\hline
$F_3$ & $c_{31} \,t $ & $\rho$ & 1 \\ 
$F_3$ & $c_{32} \,t $ & $\omega$ & 2 \\ 
\hline
\multicolumn{4}{|c|}{ Cut  amplitudes }\\
\hline
$F_1$ & $c_{14} \,   \exp[d_4t]$ & $\rho$ & 4 \\ 
$F_1$ & $c_{15} \,   \exp[d_5t]$ & $\omega$ & 5 \\ 
$F_1$ & $c_{16}\,   \exp[d_6t]$  & $b_1$ & 6 \\ 
\hline
$F_2$ & $c_{24} \, t\,  \exp[d_4t]$ & $\rho$ & 4 \\ 
$F_2$ & $c_{25} \, t\, \exp[d_5t]$ & $\omega$ & 5 \\ 
$F_2$ & $c_{26}\,  t\,  \exp[d_6t]$  & $b_1$ & 6 \\ 
\hline
$F_3$ & $c_{34}\, t\, \exp[d_4t]$ & $\rho$ & 4\\ 
$F_3$ & $c_{35} \,  t\,  \exp[d_5t]$ & $\omega$ & 5 \\ 
$F_3$ & $c_{36}\,  t\,  \exp[d_6t]$  & $b_1$ & 6 \\ 
\hline
\end{tabular}
\end{center}
\end{table}

In defining the Regge cut amplitudes we use the following parameterization based
on the absorption model~\cite{Collins2,Kellett,White1,White2,Henyey}
\begin{eqnarray}
F(s,t){=} \frac{\pi \, \beta (t)}{\log{(s/s_0)}}
 \frac{1 {+}{\cal S}\exp[{-}i\pi \alpha_c (t)]}{\sin[\pi
\alpha_c(t)]\, \Gamma[\alpha_c(t)]}\!\!
\left[\frac{s}{s_0}\right]^{\alpha_c(t)-1}\!\!\!\!\!\!,
\label{eq:trajcut}
\end{eqnarray}
with the trajectories defined by
\begin{eqnarray}
\alpha_c =\alpha_0 +\frac{\alpha^\prime\alpha_P^\prime \, t}
{\alpha^\prime + \alpha_P^\prime}\,,
\label{traj2}
\end{eqnarray}
where $\alpha_0$ and $\alpha^\prime$ are taken from the pole trajectory given
by Eqs.~(\ref{eq:traj}) and (\ref{traj1}), and
$\alpha_P^\prime{=}0.2$~GeV$^{-2}$ is the slope of the Pomeron trajectory. The
residue functions $\beta(t)$ of Eq.~(\ref{eq:trajcut}) are given in 
Table~\ref{tab0}.

In the very forward direction of the $\gamma{p}{\to}\pi^0{p}$ reaction there is
an interference of the $F_1$ amplitude with the one-photon exchange amplitude,
which is known as Primakoff effect~\cite{Primakoff}. This effect allows to determine 
the radiative decay width of the $\pi^0$-meson. However, the experimental resolution
of the $\gamma{p}{\to}\pi^0{p}$ data available presently is insufficient to
resolve the one-photon exchange amplitude. Thus, we omitted the interference
region, {\it i. e.} $|t|{<}0.04$ GeV$^2$ from the fit in order to fix the $F_1$
amplitude. But we add the one-photon exchange amplitude lateron and compare the
results with the $\gamma{p}{\to}\pi^0{p}$ differential cross sections 
available at very forward direction.

The relations between the observables analyzed in our study and the $t$-channel
helicity amplitudes are summarized in Appendix~A. The relation between the $F_i$, 
the $s$-channel helicity amplitudes and the invariant amplitudes are given in 
Appendix~B.

\section{Parameters of the model}
The resulting parameters of the model are listed in Table~\ref{tabp}. 
The achieved $\chi^2{/}dof$ amounts to 1.4. 
We find that there are some inconsistencies between data from different 
experiments. Thus, it is not possible to improve the confidence level of 
our global analysis unless these inconsistent data are removed from the 
data base. However, it is difficult to specify sensible criteria for 
pruning the data base. 

\begin{table}[t]
\begin{center}
\caption{Parameters of the model. Here $c_ {ij}$ is the coupling constant
for the $i$-th amplitude and the type $j$ of exchange, $d_j$ is a cut-off
parameter for the Regge cut amplitude.}
\renewcommand{\arraystretch}{1.3}
\label{tabp}
\begin{tabular}{|l|c|c|c|c|}
\hline
$j$ & \multicolumn{3}{|c|}{$c_{ij}$} & $d_j$\\
 & $i{=}1$ & $i{=}2$ & $i{=}3$ & \\
\hline
1 &$-12.0$ & -- & $-15.9$ & -- \\
2 & $41.4$ & - &$-56.1$ & - \\
3 &-        & $8.1$ &-- & -- \\
4 & $-2572$ & $1174$ & $-6278$ & $3.59$ \\
5 &$-18.9$  & $-4.8$ & $37.5$ & $0.64$ \\
6 &$3055$   & $-1403$ & $7526$ & $3.65$ \\
\hline
\end{tabular}
\end{center}
\end{table}

The intercept of the $\omega$-trajectory at $t{=}0$ obtained from the fit is
$\alpha_\omega$=0.641$\pm$0.003, which indeed differs from that obtained for the 
$\rho$-trajectory. The coupling constants listed in Table~\ref{tabp} show
that in case of the cut amplitudes there is some compensation between the
$\rho$, $\omega$ and $b_1$-exchange contributions. However, tiny differences 
in the trajectories are reflected in very different couplings for the
cut amplitudes. Furthermore, we find that the solution is very sensitive to the
differential cross section in the vicinity of the dip. Indeed the dip
structure results from the pole amplitudes. Since there are many
data available around the dip, {\it i.e.} at $t{\simeq}$-0.5 GeV$^2$, the
parameters are well constrained by these data and the solution turns out 
to be stable.

In order to avoid any dependence of the fit on the starting values 
of the parameters we have used the random walk method to construct the 
initial parameter vector and we have repeated the minimization procedure. 
This allows us to practically exclude that we obtain just a local minimum. 
Furthermore, an additional examination is has been done by exploring 
the results for the parameters correlation matrix in 
order to find out how unique the found minimum is.

\begin{figure}[t]
\vspace*{-6mm}
\centerline{\hspace*{4mm}\psfig{file=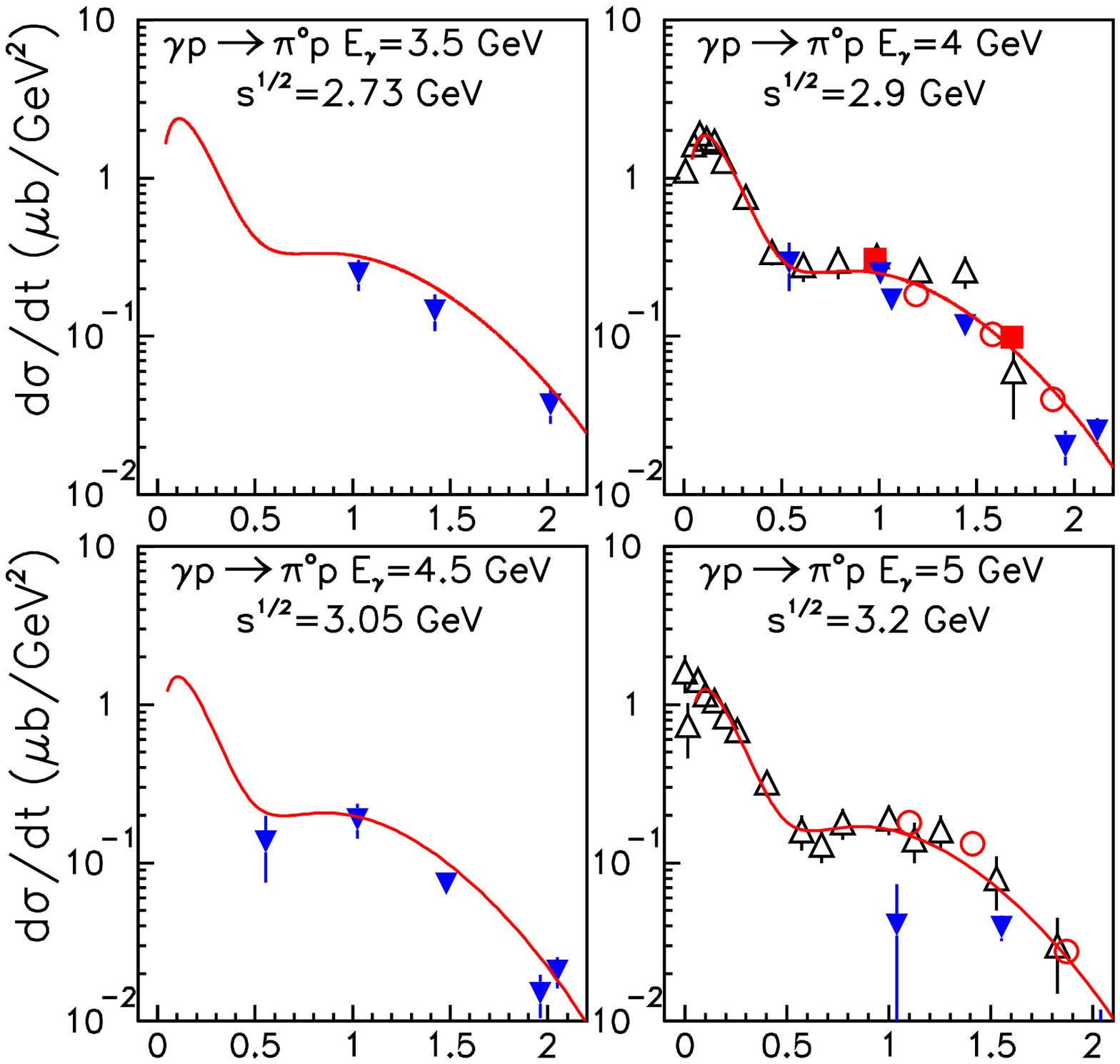,width=10cm,height=9cm}}
\vspace*{-17mm}
\centerline{\hspace*{4mm}\psfig{file=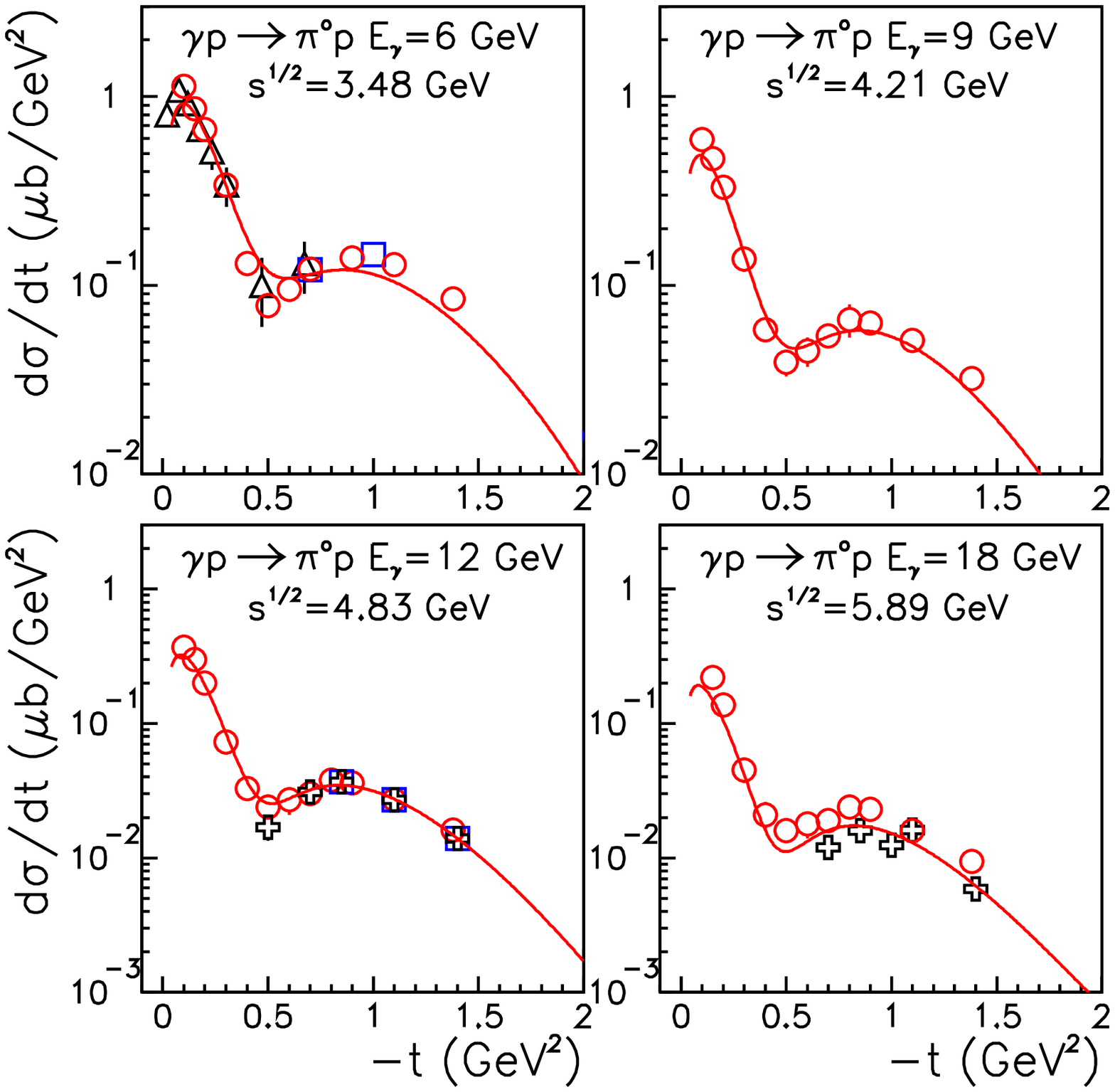,width=10cm,height=9cm}}
\vspace*{-5mm}
\caption{Differential cross section for $\gamma{p}{\to}\pi^0{p}$
as a function
of $-t$ at different photon energies $E_\gamma$ or invariant collision energies
$\sqrt{s}$. The data are taken from Refs. \cite{Bolon} (filled inverse
triangles), \cite{Braunschweig2} (open triangles), \cite{Shupe1,Shupe2}
(filled squares), \cite{Anderson3,Anderson1} (open circles), \cite{Barish} (open
squares) and \cite{Durham} (crosses). The solid lines show the results of our
model calculation.}
\label{gpine7}
\end{figure}

\section{Results of the fit}
Our results for the $\gamma{p}{\to}\pi^0{p}$ differential cross sections at
photon energies above 3~GeV are presented in Fig.~\ref{gpine7}. The model
reproduces the data quite well. As we discussed previously the data indeed
suggest a minimum or shoulder around the value $t=-0.5$~GeV$^2$, which 
was not observed
in the differential cross sections for the $\gamma{p}{\to}\pi^+{n}$ and
$\gamma{n}{\to}\pi^-{p}$ reactions. Furthermore, the dip becomes more 
pronounced with increasing photon energy. 

\begin{figure}[t]
\vspace*{-6mm}
\centerline{\hspace*{2mm}\psfig{file=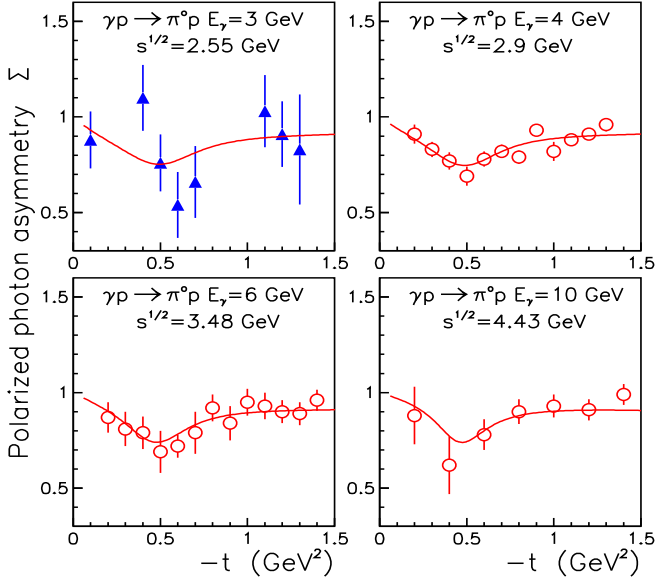,width=10cm,height=9cm}}
\vspace*{-5mm}
\caption{Polarized photon asymmetry for $\gamma{p}{\to}\pi^0{p}$
as a function of $-t$ at different photon energies $E_\gamma$ or invariant
collision energies $\sqrt{s}$. The data are taken from Refs. \cite{Bellenger}
(filled triangles) and \cite{Anderson3,Anderson1} (open circles). The solid
lines show the results of our model calculation.}
\label{gpine8}
\end{figure}

In 
Fig.~\ref{gpine8} we display the data on the polarized photon asymmetry 
available for the reaction $\gamma{p}{\to}\pi^0{p}$ at photon energies 
above 3 GeV. The data~\cite{Anderson3,Anderson1} 
at the energies $E_\gamma$=4, 6, and 10~GeV are
sufficiently precise and clearly indicate a dip around $t{\simeq}$-0.5
GeV$^2$. The Regge calculations reproduce the experimental results 
resonably well.

\begin{figure}[t]
\vspace*{-3mm}
\centerline{\hspace*{4mm}\psfig{file=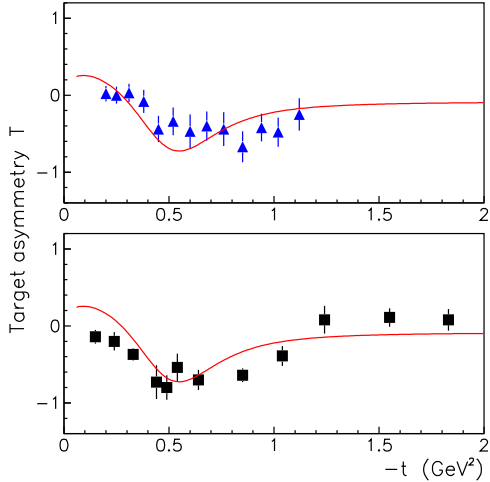,width=7.4cm}}
\vspace*{-4mm}
\caption{Target asymmetry for $\gamma{p}{\to}\pi^0{p}$
as a function of $-t$ at photon energy  $E_\gamma$=4 GeV or invariant 
collision energy $\sqrt{s}$=2.9 GeV. The triangles are data from
Ref.~\cite{Bienlein}, while the squares are from Ref.~\cite{Booth}.
The solid lines show the results of our model calculation.}
\label{gpine10b}
\end{figure}

Next we take a look at the data available for the target ($T$) and recoil ($P$)
asymmetries.
Fig.~\ref{gpine10b} shows experimental results on the target asymmetry of the 
$\gamma{p}{\to}\pi^0{p}$ reaction at a photon energy of 4 GeV.
The data~\cite{Booth,Bienlein} at the same energy are from independent measurements. 
The solid lines are the result of the Regge calculations. They are in agreement
with the data within experimental uncertainties. 

\begin{figure}[t]
\vspace*{-3mm}
\centerline{\hspace*{4mm}\psfig{file=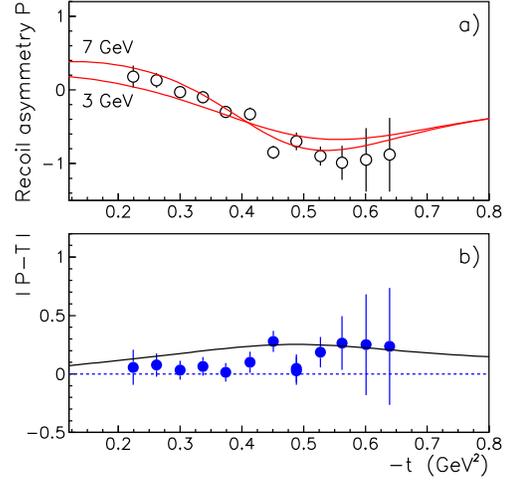,width=7.4cm}}
\vspace*{-4mm}
\caption{a) Recoil asymmetry for $\gamma{p}{\to}\pi^0{p}$
as a function of $-t$. Open circles are the data~\cite{Deutsch} obtained at
photon energies from 3 to 7 GeV. The lines show the Regge calculations at
photon energies 3 and 7 GeV. b) Illustration of the Worden inequality given by
Eq.~(\ref{ine1}). Closed circles are the difference between $P$ taken from
experiment~\cite{Deutsch} and $P$ given by the Regge calculation. 
The solid line is
$1-\Sigma$ with the polarized photon asymmetry taken from the calculations. The
dashed line indicates the case $|P{-}T|{=}0$.}
\label{gpine10a}
\end{figure}

The
open circles in Fig.~\ref{gpine10a}a) are experimental results for the 
recoil asymmetry in the reaction $\gamma{p}{\to}\pi^0{p}$ for
incident photon energies between 3 and 7 GeV~\cite{Deutsch}. 
These data allow us to examine whether the target and recoil asymmetries are 
different. This issue was
considered in Ref.~\cite{Deutsch} via a direct comparison of experimental
results~\cite{Booth,Bienlein,Deutsch} available for the target and recoil
asymmetries. It was argued that $T{\ne}P$, so that there must be higher order
contributions to the $\gamma{p}{\to}\pi^0{p}$ reaction. Indeed following
Eqs.~(\ref{obs3}) and (\ref{obs4}) the amplitude $F_4$ is not negligible
in such a case. 

However, as remarked in Ref.~\cite{Deutsch}, the measurements of the
target and recoil asymmetries were done at different energies and $P$ is
averaged over photon energies ranging from 3 to 7 GeV. In particular, 
it was emphasized that without real calculations any conclusions remain 
quite speculative. 
Now we can investigate this issue in more detail. The two lines in 
Fig.~\ref{gpine10a}a) show the Regge results for $E_\gamma$=3 GeV and
$E_\gamma$=7 GeV. Indeed the asymmetries $P$ obtained at the two energies are
slightly different. The calculations reproduce the data fairly well 
and from that we
conclude that there are no solid arguments to claim that $T{\ne}P$ at high
energies and thus to speculate about any significance of the $F_4$ amplitude.

A further examination of $F_4$ can by done by applying the
Worden inequality~\cite{Worden} given by 
\begin{eqnarray}
|P-T| \le 1-\Sigma 
\label{ine1} 
\end{eqnarray}
and shown in the Fig.~\ref{gpine10a}b). Here the closed circles indicate the
difference between the experimental results~\cite{Deutsch} for $P$ and the
calculation for $T$. The solid line is the difference $1-\Sigma$ with the 
polarized photon asymmetry taken from the calculation. Note that in this case
the Regge results are in good agreement with the data on the $T$ and $\Sigma$
asymmetries. Fig.~\ref{gpine10a} illustrates that the inequality~\cite{Worden} 
given by Eq.~(\ref{ine1}) is satisfied within the experimental uncertainties. 

\section{Predictions at energies below 3 GeV}
In this section we compare our predictions with older data for 
energies below 3 GeV but also with the most recent experimental
results~\cite{Dugger,Bartholomy} for differential cross sections collected by the
CLAS Collaboration at JLab and by CB-ELSA in Bonn. Both latter experiments cover 
the energy range up to $E_\gamma{\simeq}3$~GeV. 
As pointed out in Ref.~\cite{Dugger} the JLab results disagree with the CB-ELSA 
measurements at forward angles. It will be interesting to inspect the observed 
discrepancy with regard to the predictions by the Regge model. 
One knows from the case of charged pion photoproduction say, that the Regge 
phenomenology works well for forward angles even down to $E_\gamma{\simeq}2$~GeV. 

As stressed in many studies~\cite{Irving,Henyey,Worden}, the Regge theory is
phenomenological in nature. There is no solid theoretical derivation that allows
us to establish explicitly the ranges of $t$ and $s$ where this formalism is
applicable. Since there are several well-known nucleon resonances~\cite{PDG} in
the energy range up to $\sqrt{s}{\simeq}2.6 $ GeV, identified in
partial wave analyses~\cite{Hoehler1,Hoehler2,Koch,Cutkovsky,Manley} of
pion-nucleon scattering, we expect that deviations of our predictions from the
data will start to show up for energies from $E_\gamma{\simeq}3 $ GeV downwards.
But it will be interesting to see whether and in which observables
such discrepancies indeed occur.

We also present results utilizing the amplitudes from the 
partial wave analysis (PWA) of the GWU 
Group~\cite{Arndt,Arndt1,Arndt2,Arndt08},
which was recently extended up to $\sqrt{s}{\simeq}$2.55~GeV \cite{Dugger}.
The results we display are based on the current solution taken from 
the interactive program SAID \cite{SAID}.
The interesting question here is whether there is an energy region 
where the PWA results are in agreement with the Regge calculations. 
This would be an indication that the PWA solution might have approached 
the high energy limit given by Regge phenomenology, at least in terms of 
the considered observables. 

\begin{figure}[t]
\vspace*{-6mm}
\centerline{\hspace*{4mm}\psfig{file=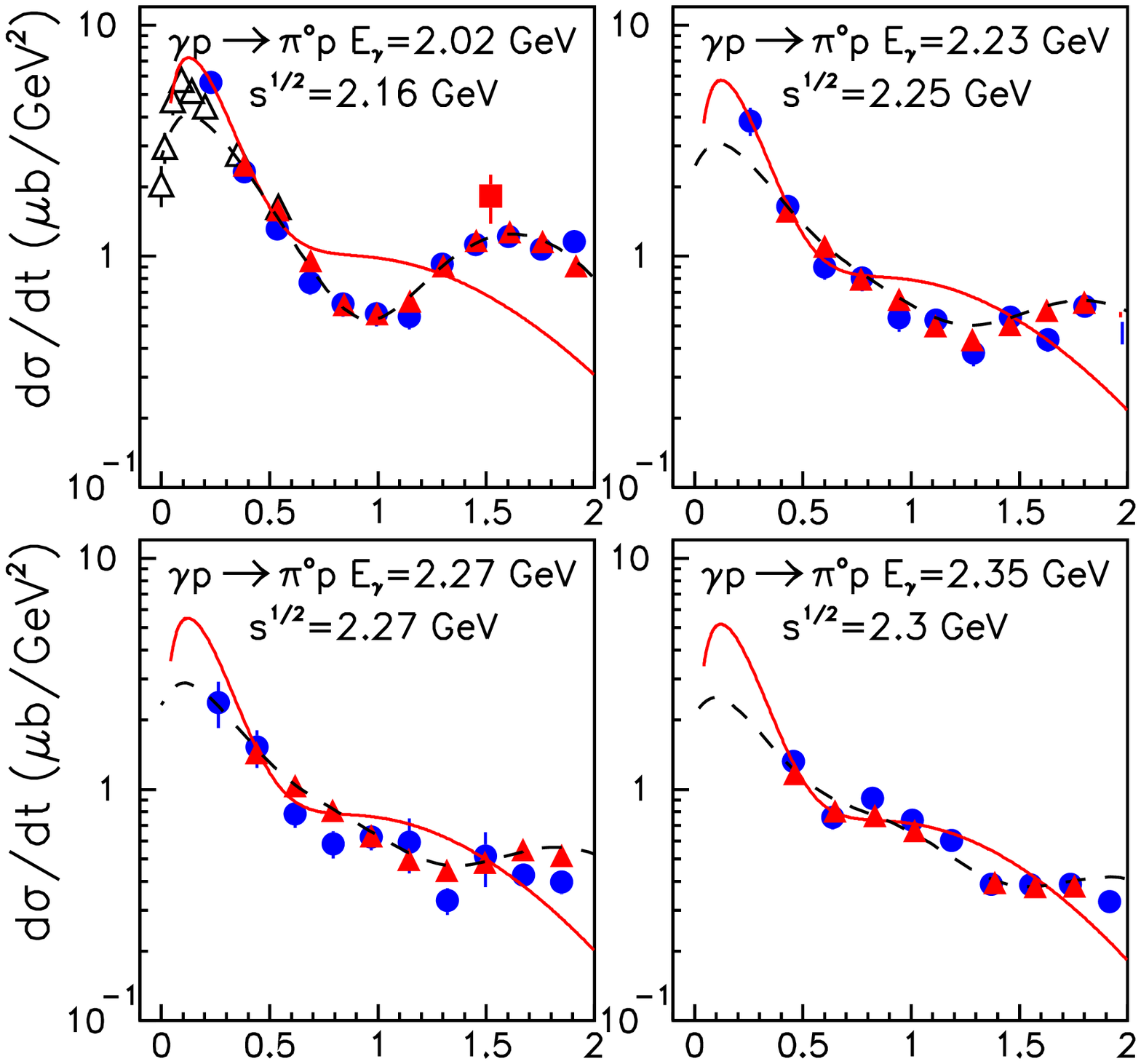,width=10.cm,height=9cm}}
\vspace*{-17mm}
\centerline{\hspace*{4mm}\psfig{file=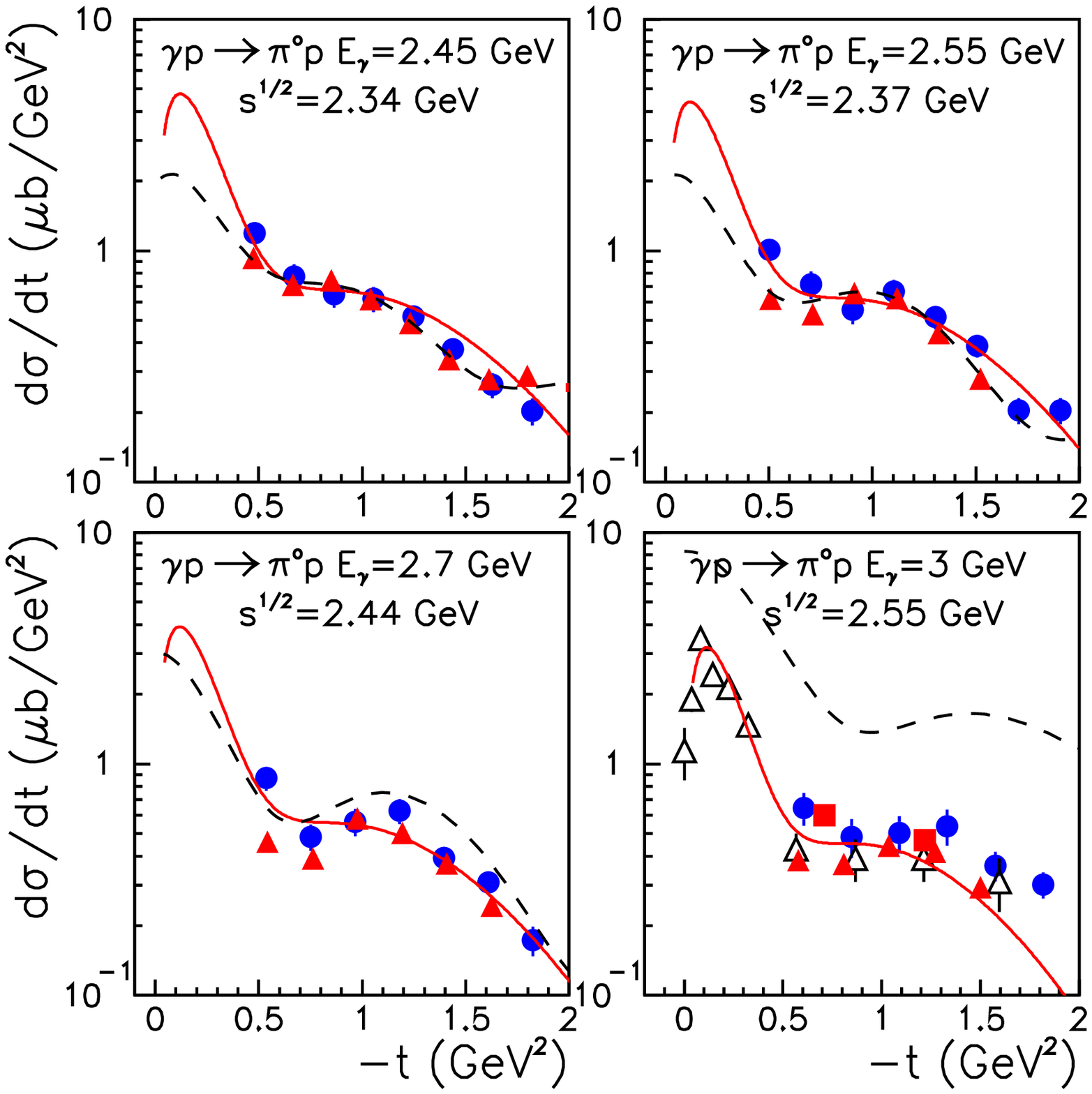,width=10.cm,height=9cm}}
\vspace*{-5mm}
\caption{Differential cross section for $\gamma{p}{\to}\pi^0{p}$ as a function
of $-t$ at different photon energies $E_\gamma$ or invariant collision energies
$\sqrt{s}$. The data are taken from Refs. \cite{Dugger} (filled triangles),
\cite{Bartholomy} (filled circles), \cite{Braunschweig2} (open triangles) and
\cite{Shupe1,Shupe2} (filled squares). The solid lines show the results of our
model calculation. The dashed lines are the results based on the GWU
PWA~\cite{SAID}.}
\label{gpine5}
\end{figure}

In Fig.~\ref{gpine5} we display differential cross sections for 
$\gamma{p}{\to}\pi^0{p}$ at photon energies from 2.02 to 3 GeV. 
This energy region corresponds to invariant collision energies of
2.16${\le}\sqrt{s}{\le}$2.55~GeV. As expected with decreasing energy
the predictions of the Regge model differ more and more from the data. 
But at least down to roughly $\sqrt{s}{=}$2.34 GeV the results are still
fairly well in line with the experimental information, considering the 
variations between the existing measurements. Note that the Regge
model reproduces the strong rise of the cross section for very small 
angles, as seen in data from Ref.~\cite{Braunschweig2} at 
$\sqrt{s}{=}$2.55 GeV, very accurately. 

\begin{figure}[t]
\vspace*{-6mm}
\centerline{\hspace*{2mm}\psfig{file=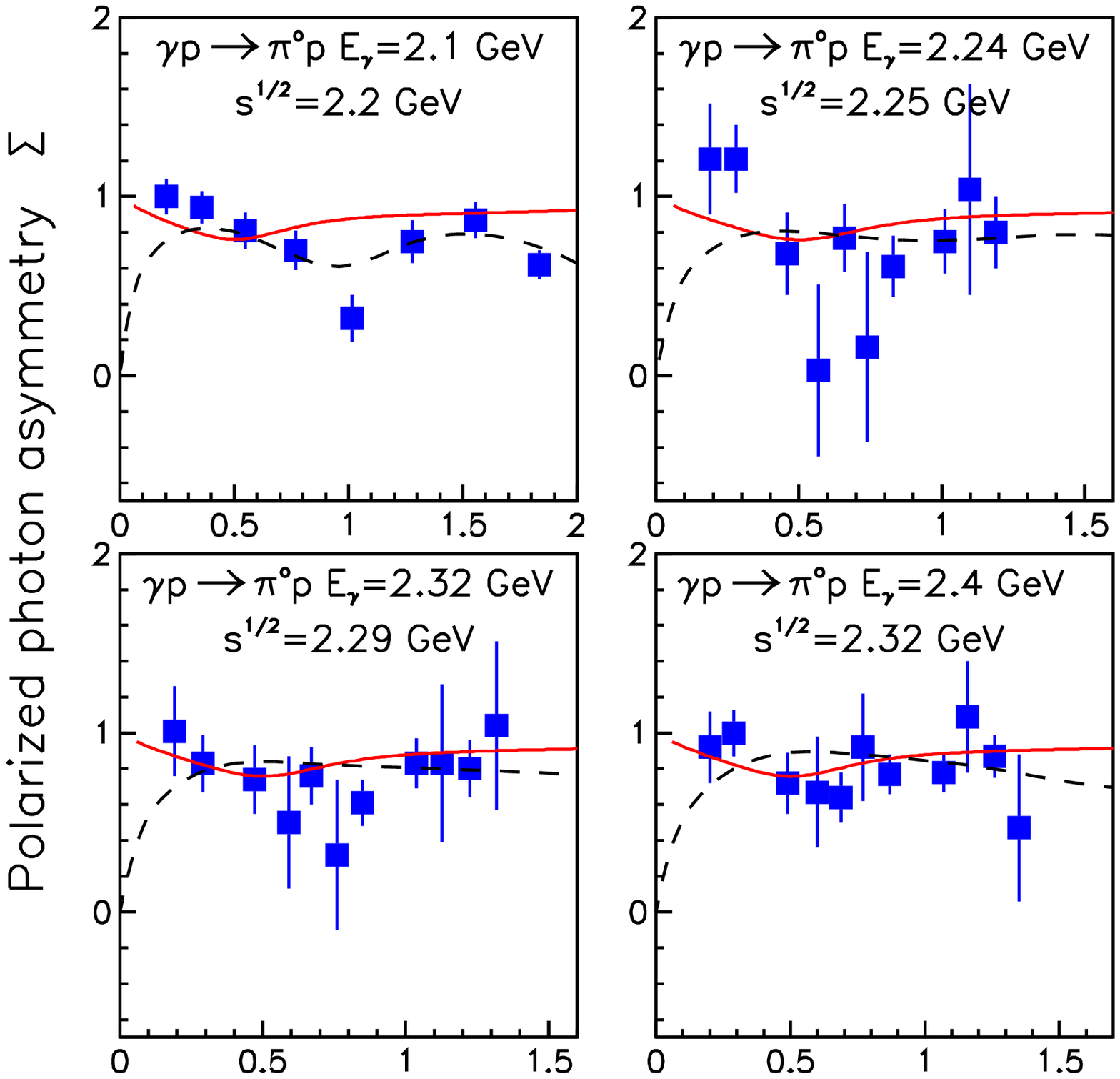,width=10.cm,height=9cm}}
\vspace*{-17mm}
\centerline{\hspace*{2mm}\psfig{file=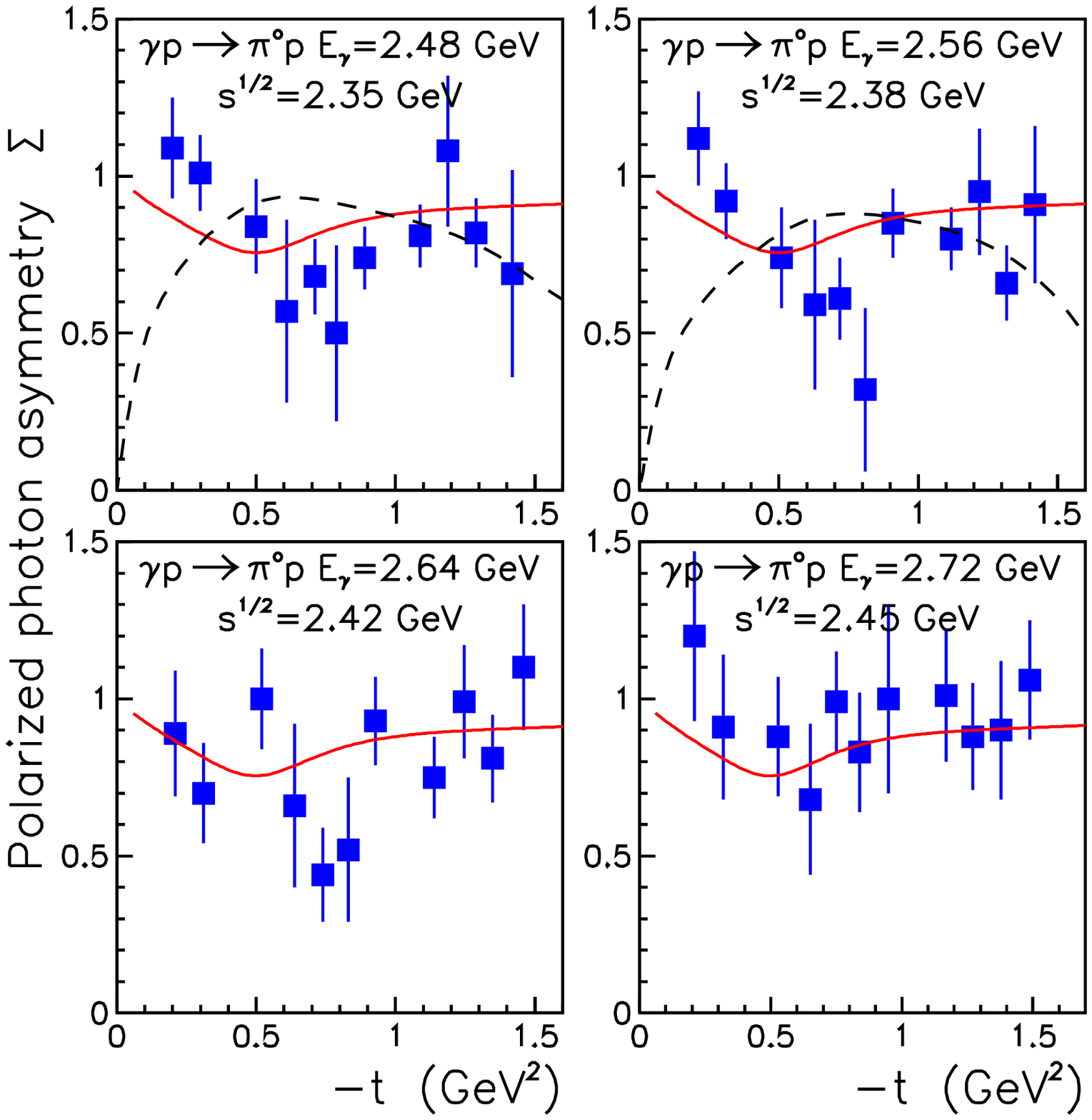,width=10.cm,height=9cm}}
\vspace*{-5mm}
\caption{Polarized photon asymmetry for $\gamma{p}{\to}\pi^0{p}$
as a function of $-t$ at different photon energies $E_\gamma$ or invariant
collision energies $\sqrt{s}$. The data are taken from Refs. \cite{Bussey}. The
solid lines show the results of our model calculation. The dashed lines are the
results based on the GWU PWA~\cite{SAID}.}
\label{gpine9}
\end{figure}

It is evident from Fig.~\ref{gpine5} that there is a systematic
disagreement between the CLAS results~\cite{Dugger} (filled triangles)
and the ELSA measurement (filled circles) at the higher energies,
especially in the forward direction. With regard to our Regge results 
there is no clear preference for any of the two measurements.
However, at least in the vicinity of the dip or shoulder our results 
are definitely more in line with the ELSA measurement.
We would like to emphasize that neither the CLAS nor the ELSA data 
were included in our fit. 

One can see from Fig.~\ref{gpine5} that the GWU 
PWA~\cite{Arndt,Arndt1,Arndt2,Arndt08}, shown by the dashed lines, 
reproduces the data from JLAB (to which 
it was fitted) very well up to photon energies of 2.55 GeV 
or invariant energies of $\sqrt{s}$=2.37 GeV. 
Also, for invariant collisions energies around 2.34$\simeq$2.37~GeV 
the GWU PWA results and the Regge calculations are similar, at least
qualitatively and for the range 0.5$\le -t \le$ 1.2 GeV$^2$. 
It is important to note that our model does not include any resonance 
contribution. The GWU PWA indicates the presence of the 
$G_{39}(2400)$ $\Delta$-resonance at the upper end of the 
fit to $\pi N$ elastic scattering data~\cite{Arndt08}. But it remains 
unclear whether this resonance also has a noticable impact on their results 
for neutral pion photoproduction. 

The polarized photon asymmetry, measured~\cite{Bussey} at photon energies 
from 2.1 to 2.75 GeV, is shown in Fig.~\ref{gpine9}. Unfortunately, the 
experimental results are afflicted by large uncertainties and, therefore, 
do not allow us to draw any more quantitative conclusions on the reliability 
of the Regge predictions. But the results are roughly in line with the
data over the whole energy region. The GWU PWA is in reasonable agreement
with the data at 2.1 GeV, but develops a qualitatively different 
behavior with increasing energy. 
The data seem to indicate the presence of a dip at around 
$t{\simeq}$-0.8 GeV$^2$ at almost all shown energies.
The Regge model produces such a dip, but near the value $t{\simeq}$-0.5~GeV$^2$, 
while the results from GWU PWA exhibit a dip structure only at the 
lowest energy considered.

\begin{figure}[t]
\vspace*{-6mm}
\centerline{\hspace*{2mm}\psfig{file=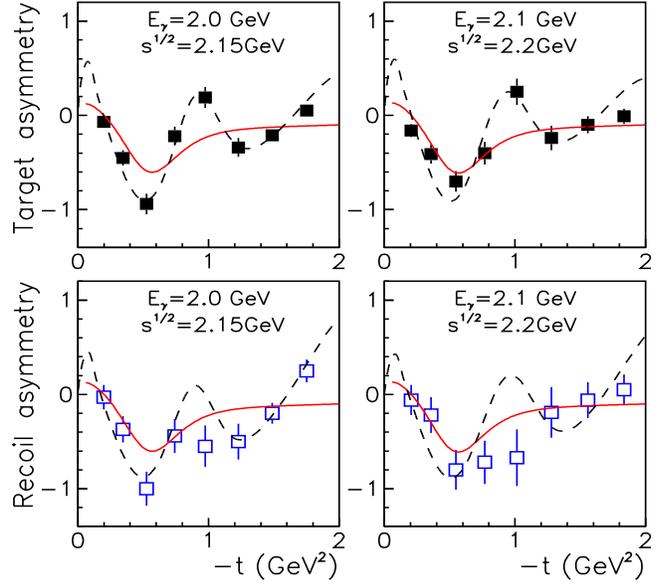,width=10.cm,height=9cm}}
\vspace*{-5mm}
\caption{Target (filled squares) and recoil (open squares) asymmetries for 
$\gamma{p}{\to}\pi^0{p}$
as a function of $-t$ at different photon energies $E_\gamma$ or invariant
collision energies $\sqrt{s}$. The data are taken from Refs. \cite{Bussey}. The
solid lines show the results of our model calculation. The dashed lines are the
results based on the GWU PWA~\cite{SAID}.}
\label{gpine10}
\end{figure}

\begin{figure}[t]
\vspace*{-6mm}
\centerline{\hspace*{3mm}\psfig{file=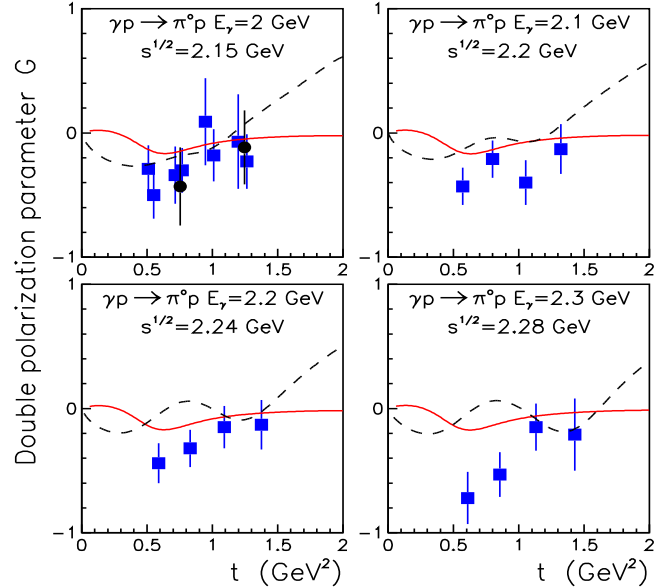,width=10.cm,height=9cm}}
\vspace*{-5mm}
\caption{Double polarization parameter $G$ for $\gamma{p}{\to}\pi^0{p}$
as a function of $-t$ given at different photon energies $E_\gamma$.
The  circles are results from Ref.~\cite{Booth}, while the squares show the data
from Ref.~\cite{Bussey1}. The
solid lines show the results of our model calculation. The dashed lines are the
results based on the GWU PWA~\cite{SAID}.}
\label{gpine1}
\end{figure}

Fig.~\ref{gpine10} shows target and recoil asymmetries in the
$\gamma{p}{\to}\pi^0{p}$ reaction, measured~ \cite{Bussey} at photon energies of
2 and 2.1 GeV. The data on the target asymmetry have quite small uncertainties 
and exhibit a significant variation with four-momentum transfer squared. 
Apparently, at these energies $T{\ne}P$. The Regge calculations reproduce 
the recoil and target asymmetry roughly, but only at very forward angles. 
On the other hand, the GWU PWA describes $T$ as well as $P$ fairly well over
the considered $t$ range. 

\begin{figure}[t]
\vspace*{-6mm}
\centerline{\hspace*{3mm}\psfig{file=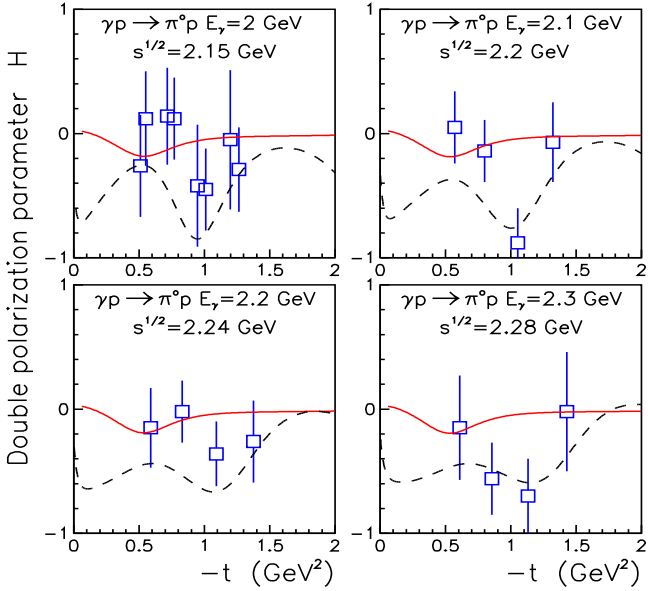,width=10.cm,height=9cm}}
\vspace*{-5mm}
\caption{Double polarization parameter $H$ for $\gamma{p}{\to}\pi^0{p}$
reaction at different photon energies $E_\gamma$. The data are from
Ref.~\cite{Bussey1}. The solid lines show the results of our model 
calculation. The dashed lines are the
results based on the GWU PWA~\cite{SAID}.}
\label{gpine2}
\end{figure}

Some comments with regard to the observed difference between $T$ and $P$
at those energies seem to be in order. In the case of $t$-channel non-resonant 
contributions the $F_4$ helicity amplitude is given by higher order
corrections and thus we neglect it, as was discussed
previously~\cite{Sibirtsev07}. But even if $F_4$ does not vanish 
its general influence on the various observables is expected~\cite{Irving,Rahnama3} 
to be small. Note that in approaches based on an effective Lagrangian, which 
are commonly used at low energies, the contribution from vector-meson exchanges 
to pseudoscalar meson photoproduction also results in a vanishing 
invariant amplitude $A_3$ and, thus, following Eq.~(\ref{invari}), one
would expect $F_4{=}0$. However, at the same time resonances can
contribute~\cite{Benmerrouche,Sibirtseve} to the amplitude $F_4$ so that
the difference between $T$ and $R$ might be explained in a natural way. 

For the $\gamma{p}{\to}\pi^0{p}$ reaction there are also data~\cite{Bussey1}
for the double polarization parameters $G$ and $H$.
These data are important for fixing the sign of the amplitude $F_2$, as is
obvious from Eqs.~(\ref{obs6}) and (\ref{obs8}). Since there are no data for
the double polarization parameters at higher energies we cannot determine
the sign of the $F_2$ amplitude within our fitting procedure.
Therefore, we decided to use the data at low energies to fix that 
ambiguity for the parameters corresponding to the $F_2$ amplitude,
listed in the Table~\ref{tab0}.

Fig.~\ref{gpine1} shows the double polarization parameter, $G$, measured at 
photon energies from 2 to 2.3 GeV, which correspond to invariant energies of 
2.15 to 2.28 GeV. The solid lines are the results obtained from our Regge
model. They are only very qualitatively in line with the data.
A similar conclusion might also be drawn when comparing the PWA results
with the measurements.

Fig.~\ref{gpine2} shows the double polarization parameter, $H$, measured 
for photon energies between 2 and 2.3 GeV. 
The solid lines are the results obtained from the Regge model. Again,
these are qualitatively in line with the experiment.
It is interesting that the PWA predicts large negative 
values for $H$ that vary strongly with the four-momentum momentum squared,
while the Regge model predicts $H{\simeq}0$.

\section{Further results and discussion}

\subsection{Total cross sections and the ratio of {\boldmath$\pi^0$}
photoproduction on neutrons and protons}
Information about the relative contributions of the isovector $\rho$ 
(and $b_1$) exchange and the isoscalar $\omega$ exchange amplitudes can 
be obtained by comparing the differential cross sections for $\pi^0$-meson 
photoproduction on neutrons and on 
protons~\cite{Irving1,Barker2,Froyland,Argyres}. For the proton target 
the total reaction amplitude is given by the sum of the isovector and
isoscalar contributions, while for the neutron target it is given by 
their difference.

\begin{figure}[t]
\vspace*{-6mm}
\centerline{\hspace*{3mm}\psfig{file=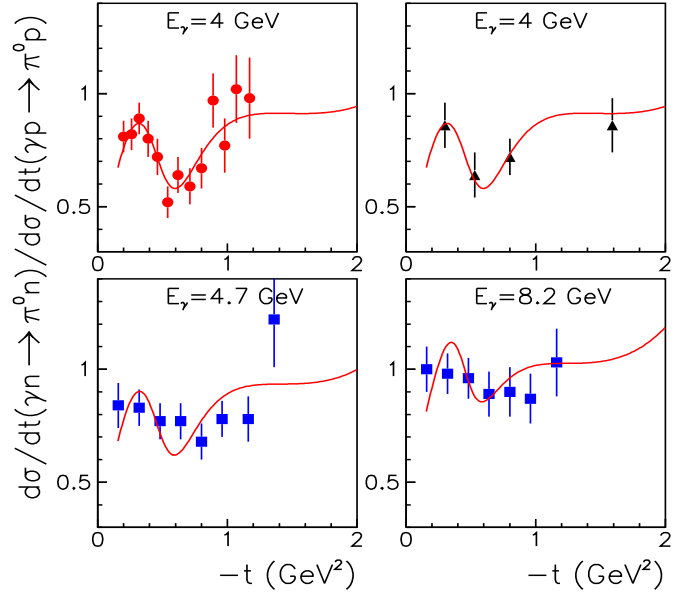,width=10.cm,height=9cm}}
\vspace*{-4mm}
\caption{The ratio of the differential cross sections 
for $\pi^0$-photoproduction on neutrons and protons as a function of 
$-t$ for different photon energies $E_\gamma$.
The data are from Refs.~\cite{Braunschweig3} (circles), \cite{Bolon4}
(triangles) and \cite{Osborne4} (squares). The lines show the results
of our model calculation.}
\label{gpine12}
\end{figure}

In 
Fig.~\ref{gpine12} we present the ratio of the differential cross sections 
for $\pi^0$-photoproduction on neutrons and protons for different 
photon energies. The available data demonstrate that the ratio $R{\ne}1$, which
contradicts a statement given in Ref.~\cite{Guidal}. Note that the precise
measurement~\cite{Braunschweig3} at $E_\gamma$=4 GeV indicates that the ratio
depends considerably on $t$. 

\begin{figure}[t]
\vspace*{-2mm}
\centerline{\hspace*{3mm}\psfig{file=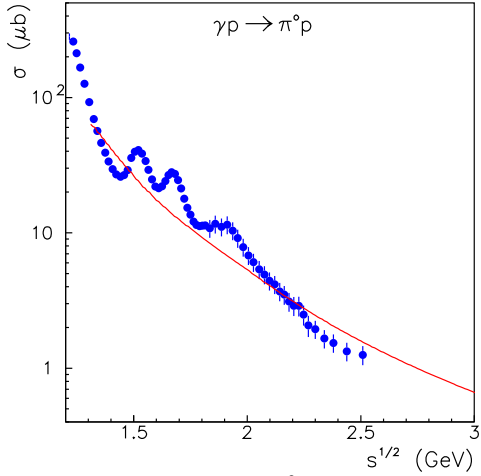,width=7.2cm}}
\vspace*{-4mm}
\caption{Total cross section for $\gamma{p}{\to}\pi^0p$. The
circles are the experimental results taken from Ref.~\cite{Bartholomy}. 
The lines show the results of our model calculation.}
\label{gpine11}
\end{figure}

Fig.~\ref{gpine11} shows the total cross section for the $\gamma{p}{\to}\pi^0p$
reaction as a function of the invariant collision energy. The experimental 
results were obtained~\cite{Bartholomy} by integration over the angular
distributions and involve an extrapolation into the forward and backward 
regions using the results from the isobar model of 
Anisovich et al.~\cite{Klempt}. The solid line is the
Regge result, obtained by integration of the calculated 
differential cross section over the range $|t|{\le}$2 GeV$^2$.

\begin{figure}[t]
\vspace*{-6mm}
\centerline{\hspace*{3mm}\psfig{file=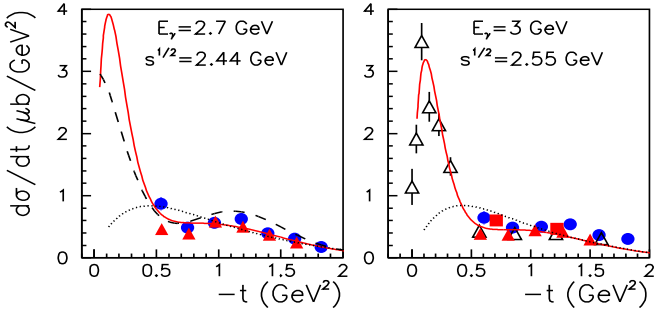,width=10.cm,height=9cm}}
\vspace*{-40mm}
\caption{Differential cross section for $\gamma{p}{\to}\pi^0{p}$ as a function
of $-t$ at $E_\gamma=$ 2.7 and 3 GeV.
The data are taken from Refs. \cite{Dugger} (filled triangles),
\cite{Bartholomy} (filled circles), \cite{Braunschweig2} (open triangles) and
\cite{Shupe1,Shupe2} (filled squares). The solid lines show the results of our
model calculation. The dashed line is the result based on the GWU 
PWA~\cite{SAID} while the dotted line is the result of the isobar model
\cite{Klempt} as given in Ref.~\cite{Bartholomy}.
}
\label{gpine5a}
\end{figure}

At first sight it looks as if the Regge model would overestimate the 
integrated cross section significantly at the higher energies, i.e. 
at energies where it is actually expected to agree 
with the data. 
However, the amplitude that is used for obtaining those cross sections 
in \cite{Bartholomy} has some shortcomings in the forward direction,
as one can see in Fig.~\ref{gpine5a}. Specifically, it fails badly  
to describe the data at very small angles \cite{Braunschweig2},
whereas the Regge model reproduces even those data rather 
well. Consequently, a determination of the total cross section that
utilizes that amplitude for extrapolating to forward angles will 
necessarily underestimate the ``real'' value. On the contrary, based on 
the quality of our fit to the small-angle data at 3~GeV, one 
expects that the predictions of the Regge model for the integrated 
cross section should be very realistic. 
It is interesting to see that the results of the isobar model 
\cite{Klempt} and of our Regge fit practically coincide in the
range $1{<}-t{<}2$ GeV$^2$. 

\begin{figure*}[t]
\vspace*{-3mm}
\centerline{\hspace*{3mm}\psfig{file=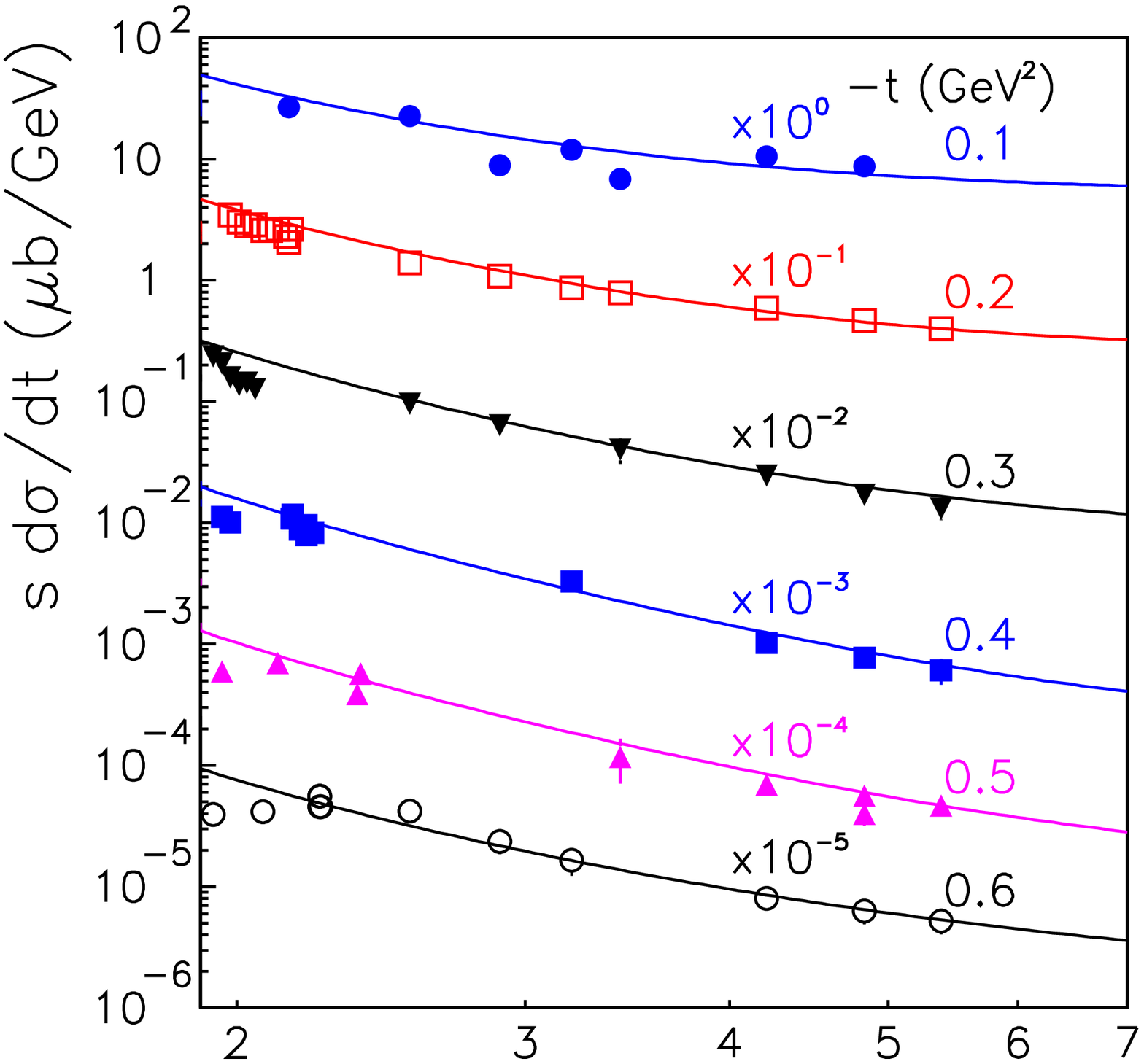,width=7cm}\hspace*{-10mm}
\psfig{file=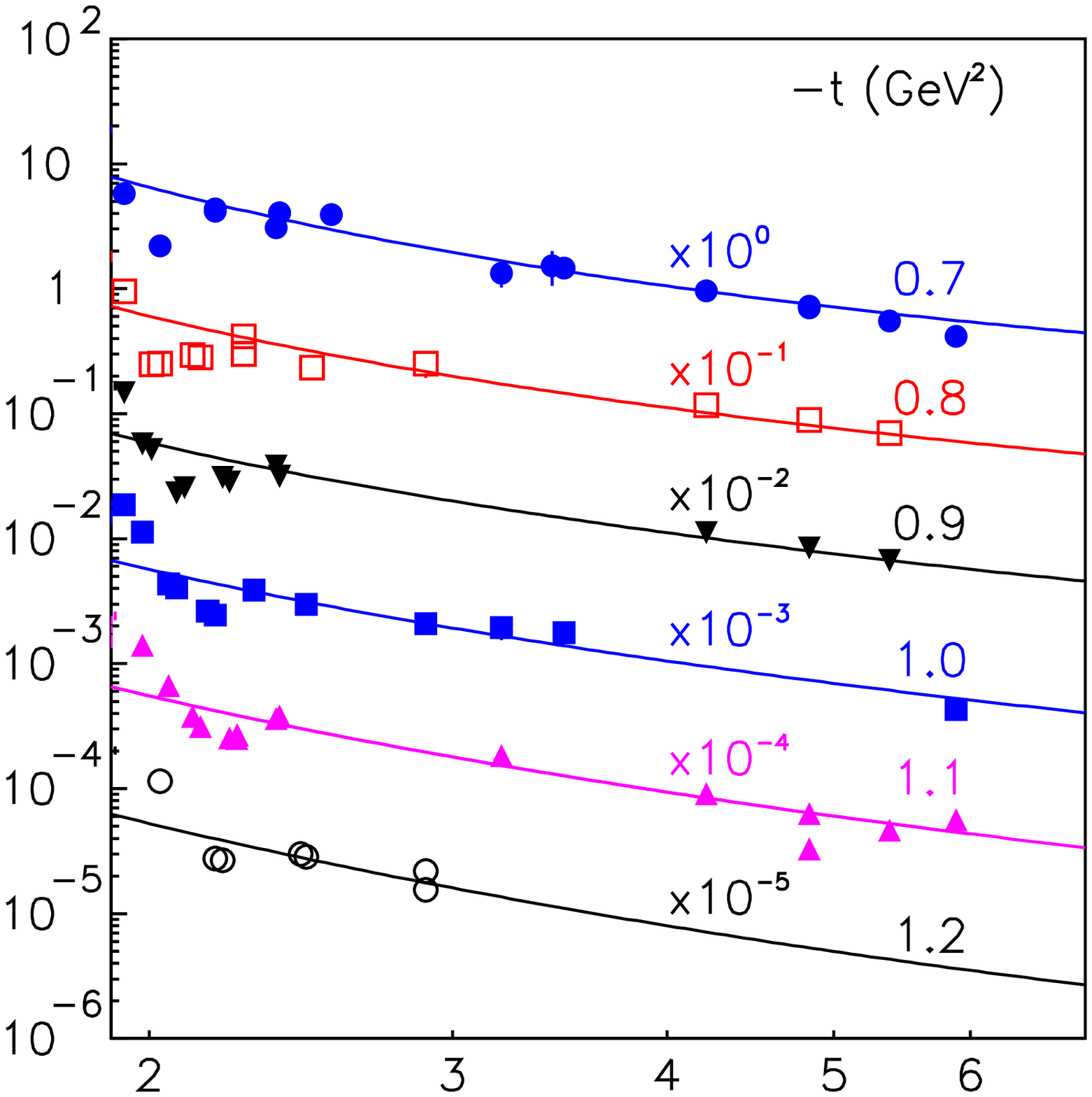,width=7cm}}
\vspace*{-10mm}
\centerline{\hspace*{3mm}\psfig{file=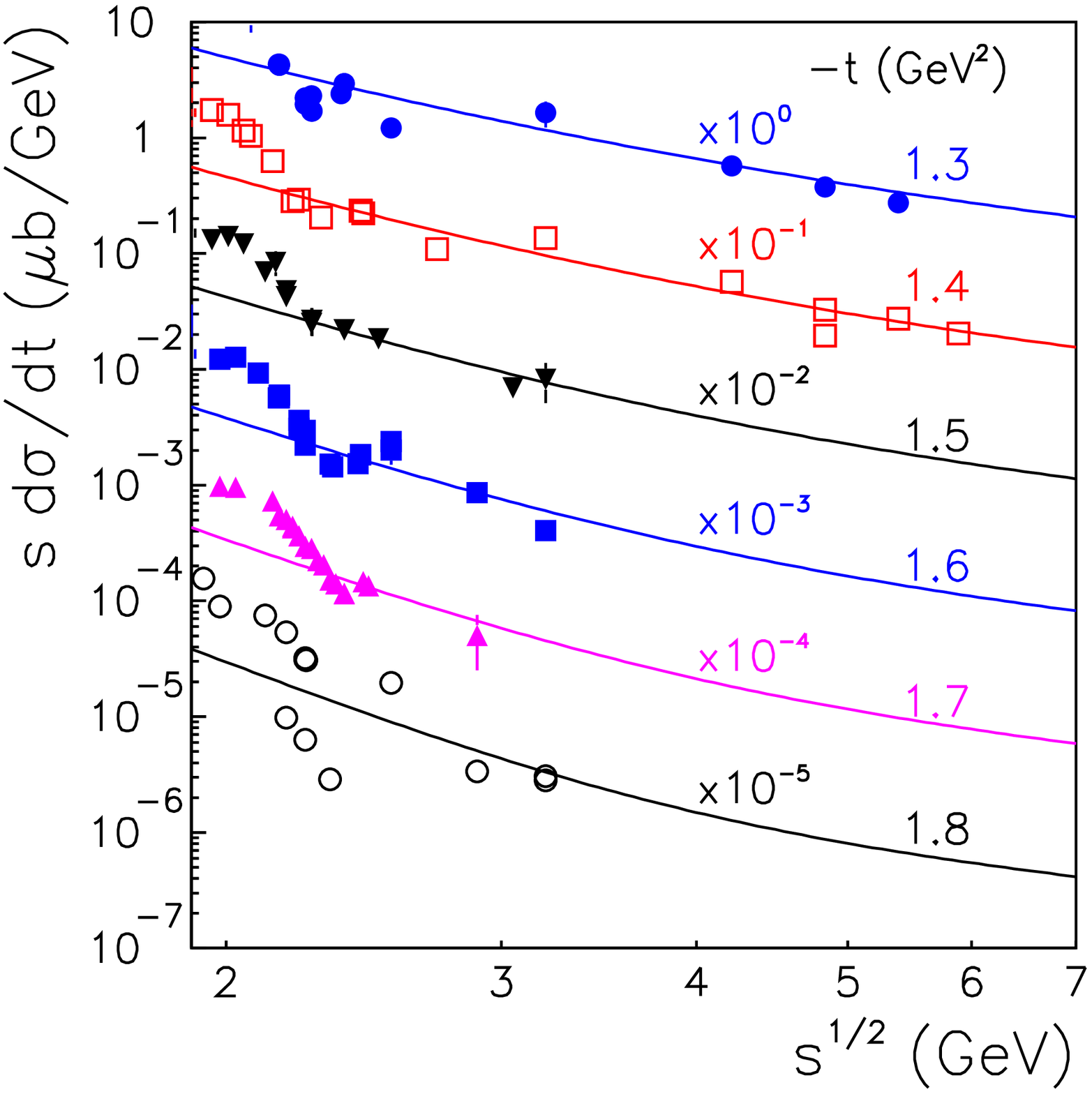,width=7cm}\hspace*{-10mm}
\psfig{file=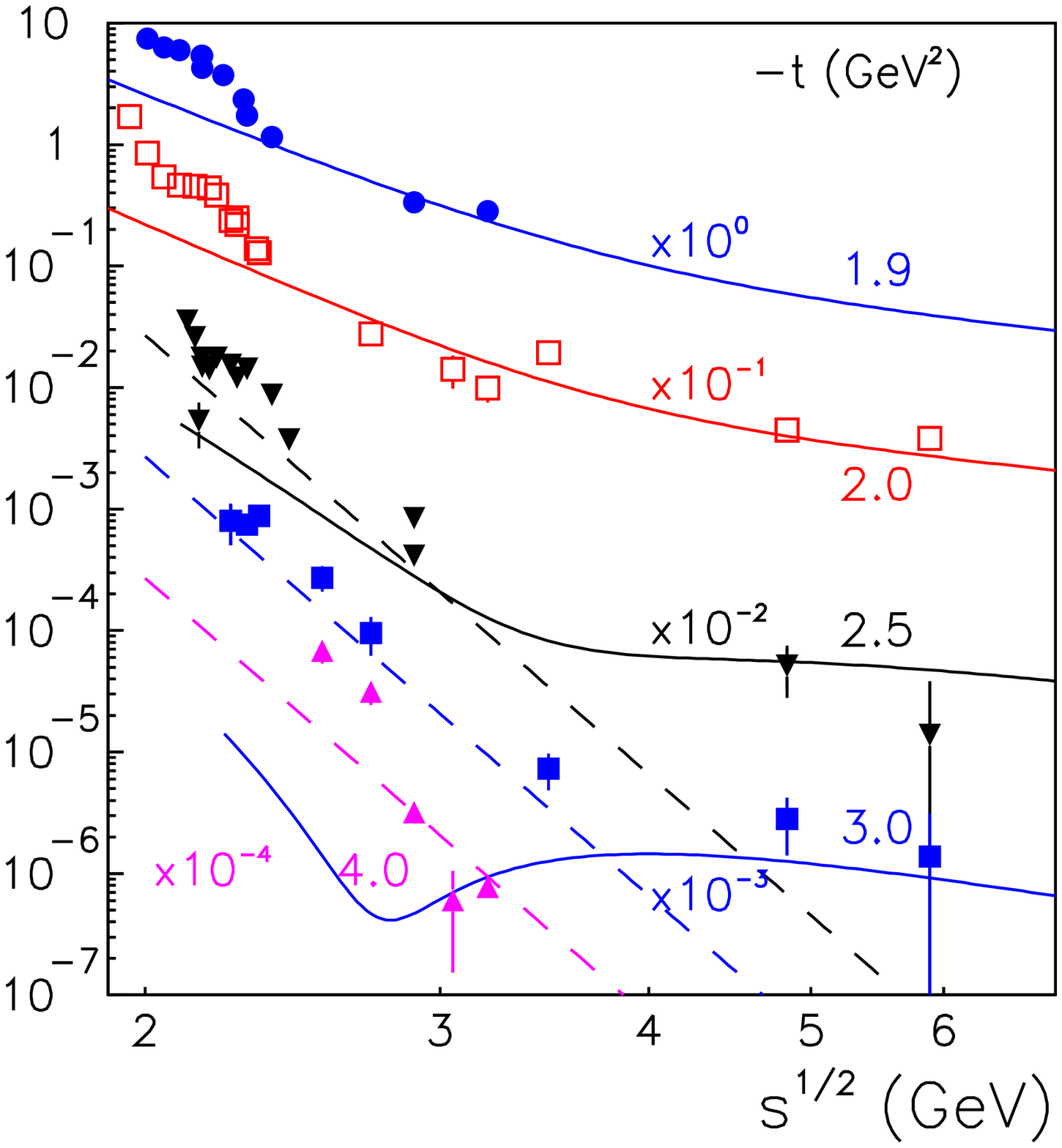,width=7cm}}
\vspace*{-1mm}
\caption{Differential cross section for $\gamma{p}{\to}\pi^0{p}$ as a function
of the invariant collision energy, at selected fixed values of $t$. The
symbols show the data considered in the present paper. The solid lines 
are the results of our model calculation. The dashed lines are the results 
obtained with Eq.~(\ref{count}). The data and lines are scaled with 
powers of 10.}
\label{fixt}
\end{figure*}

\subsection{Results for fixed four-momentum transfer}
To complete our analysis of the data we take a look at the
energy dependence of the $\gamma p{\to}\pi^0p$ differential cross sections 
at fixed four-momentum transfer squared $t$. 
This allows us to shed light on the applicability of the Regge phenomenology 
with respect to the $s$- as well as the $t$ dependence. 
It also facilitates the inspection as to whether potential discrepancies 
between the calculations and data exhibit any systematic features. 

Fig.~\ref{fixt} shows the data considered in the present analysis. Here the
differential cross sections are multiplied by the squared invariant collision
energy $s$. We multiply with this factor because the high energy limit of the
cross section at small $-t$, as given by the Regge formalism, is proportional 
to $1/s$.
Therefore, at high energies and small $-t$ we expect that $s d\sigma/dt$
approaches a constant value. For ease of comparison we scale the data 
and the curves by powers of 10. 

The solid lines in Fig.~\ref{fixt} are the results of our Regge model.
At very small $|t|$ the Regge model reproduces the data rather well, even 
down to energies of $\sqrt{s}{\simeq}2$ GeV. 
In general the data seem to agree with the high energy limit as given by 
the Regge phenomenology from energies of $\sqrt{s}=$2.5 - 3 GeV upwards, 
at least for the range $0.5{\le}|t|{\le}2$ GeV$^2$.
Below this energy region the data show sizeable variations with 
regard to the predictions of the Regge model. Specifically, for low
energies and larger $|t|$ our model results deviate systematically from 
the data and the discrepancy increases with increasing squared four-momentum 
transfer.

Furthermore, as the squared four-momentum increases the energy dependence of the
data and the Regge calculations becomes steeper. 
Although we fit the data in 
the range $t{\ge}-$2 GeV, the calculations still describe the experimental results 
at $t{=}-$2.5 GeV fairly well. However, the Regge model substantially 
underestimates all data at larger $|t|$. As we showed in Ref.~\cite{Sibirtsev07}, 
the data on charged pion photoproduction at energies $\sqrt{s}{\ge}2.7$ GeV 
and at large four-momentum transfer squared are practically independent of $t$ 
and are in line with the Dimensional
Counting Rule~\cite{Brodsky,Matveev}. According to the DCR for the
invariant amplitude $M$ the energy dependence of the differential
cross-section is given as
\begin{eqnarray}
\!\!\!\!\!\!\frac{d\sigma}{dt}{=}\frac{|M|^2 F(t)}{16\pi(s -m_N^2)^2}
{=} \frac{c s^{-(n_i-2)−(n_f-2)}F(t)}{16\pi s^2}{\propto} c s^{-7}F(t),
\label{count}
\end{eqnarray}
in the limit $m_N{\ll}s$, where $m_N$ is the mass of the nucleon.
Here $c$ is a normalization constant, while $n_i$ and $n_f$ are the total number
of elementary fields in the initial and final states, respectively. For single
pion photoproduction $n_i$=4 and $n_f$=5. 
Furthermore, $F(t)$ is a form-factor, which does not depend on the energy $s$ 
but accounts for the $t$ dependence of the hadronic wave functions and partonic
scattering. We found that, within the experimental uncertainties, the data
on $\pi^+$ and $\pi^-$-meson photoproduction indicate that $F(t)$ is almost
constant, {\it i.e.} does not depend on the squared four momentum. Moreover, it
turns out that both negative and positive pion photoproduction can be described
with the same normalization $c$=11 mb$\cdot$GeV$^{12}$, when assuming
that $F(t){=}$1.

The dashed lines in Fig.~\ref{fixt} show the results obtained using
Eq.~(\ref{count}) with
the normalization constant given above. Apart from two
experimental points at $\sqrt{s}{\ge}$5 GeV the data are in good agreement with
the ansatz based on the DCR. However, note that at least at $t{=}-$2.5 GeV$^2$ 
the Regge calculation reproduces the data better than the DCR.

\section{Primakoff effect}

\begin{figure}[t]
\vspace*{-4mm}
\centerline{\hspace*{3mm}\psfig{file=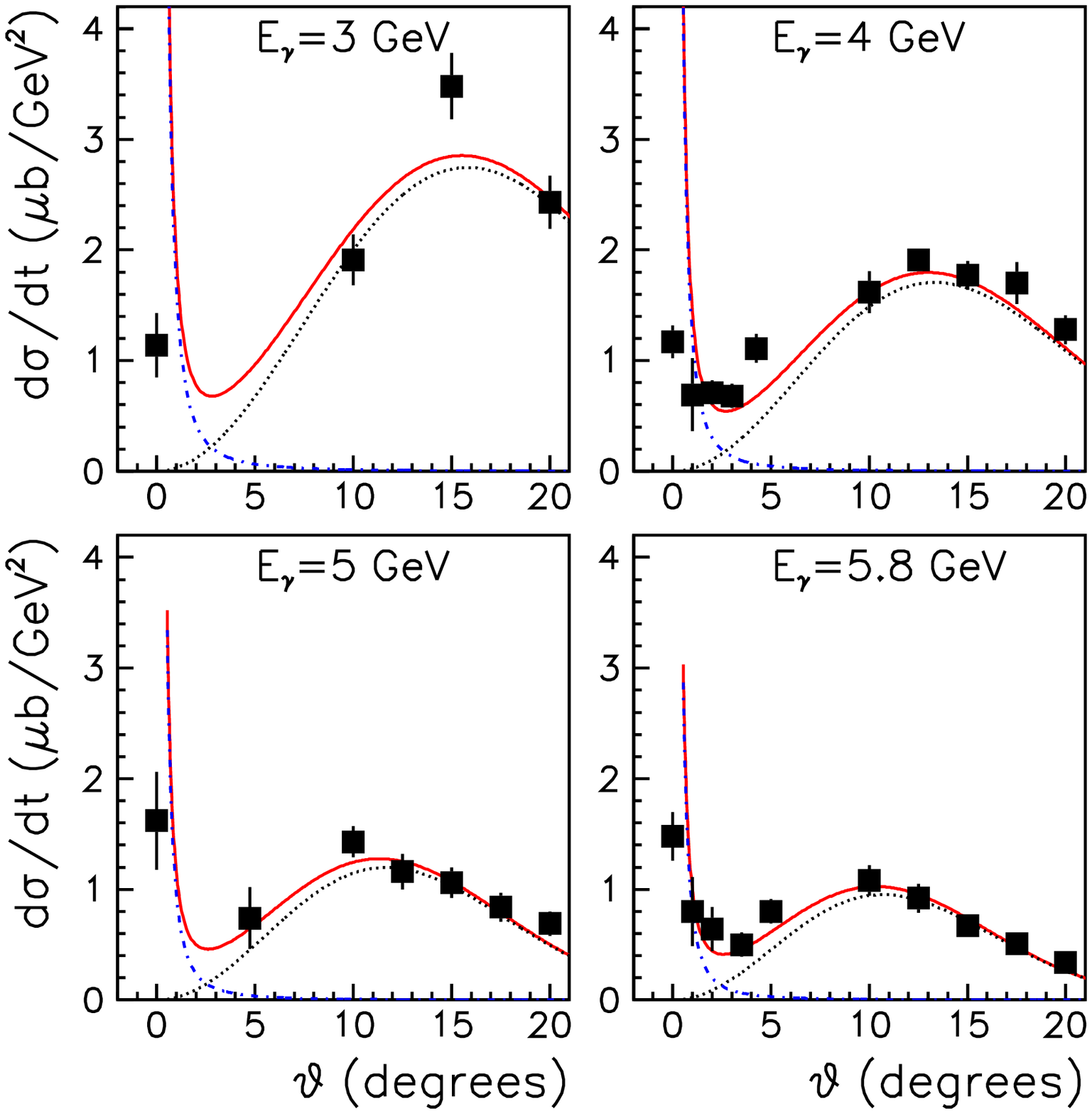,width=10.cm,height=9cm}}
\vspace*{-5mm}
\caption{Differential cross section for $\gamma{p}{\to}\pi^0{p}$ as a function
of the angle $\theta$ in the cm system shown for different photon energies
$E_\gamma$. The data are taken from Ref.~\cite{Braunschweig2}. The dotted lines
show the Regge calculations without one photon exchange, while the solid 
lines are the results obtained with inclusion of the one photon exchange. 
The dash-dotted lines are results for the one photon exchange alone. 
}
\label{prima1}
\end{figure}

\begin{figure}
\vspace*{-3mm}
\centerline{\hspace*{3mm}\psfig{file=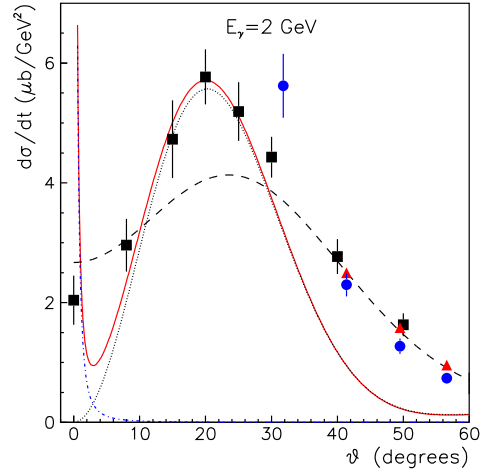,width=7.2cm}}
\vspace*{-4mm}
\caption{Differential cross section for $\gamma{p}{\to}\pi^0{p}$ as a function
of the angle $\theta$ in the cm system shown for photon energy $E_\gamma$=2 GeV.
The data are taken from Refs.~\cite{Braunschweig2} (squares), \cite{Dugger}
(triangles) and \cite{Bartholomy} (circles). The dotted line shows the
Regge calculations without one photon exchange, while the solid line is the 
results obtained with inclusion of one photon exchange. The dash-dotted 
line is the result for one photon exchange alone. 
The dashed line indicates the results based on the GWU PWA \cite{SAID}.
}
\label{prima2}
\end{figure}

The Primakoff effect \cite{Primakoff} has not only been observed
in neutral pion photoproduction on nuclei but also in the 
$\gamma{p}{\to}\pi^0{p}$ reaction \cite{Braunschweig2}.
This effect dominates the reaction cross section at 
low momentum transfer and can be used for the determination 
of the $\pi^0{\to}\gamma\gamma$ decay width. But there is an 
interference of the $F_1$ amplitude with the Primakoff 
(one photon exchange) amplitude. 
Thus, the determination of the $\pi^0$ radiative decay requires a precise 
knowledge of the hadronic part of the amplitude as well as accurate data.
Unfortunately, the available experimental results~\cite{Braunschweig2} on the 
differential cross section in the near forward direction are afflicted by considerable
uncertainties, as is illustrated in Fig.~\ref{prima1}. Here the dotted lines
show the result of our Regge model without one photon exchange, which 
reproduces the data at angles above 10$^\circ$, say, rather well.

The solid lines in Fig.~\ref{prima1} are results obtained with the Primakoff 
amplitude included, where the latter is given by
\begin{eqnarray}
F^P = \frac{8 m_p}{t} \sqrt{\frac{\pi \Gamma(\pi^0{\to}\gamma\gamma)}{m_\pi^3}}
F_D(t) = \sqrt{\Gamma} \hat F^P(t)\ .
\label{prima}
\end{eqnarray}
Here $m_p$ and $m_\pi$ are the proton and pion mass, respectively,
$\Gamma$ is the $\pi^0{\to}\gamma\gamma$ decay width and $F_D$ is
the Dirac form factor of the proton. For the latter we adopt 
the parameterization given in Ref.~\cite{Donnachie},
\begin{eqnarray}
F_D (t) = \frac{4m_p^2 - 2.8t}{4m_p^2-t}\frac{1}{(1-t/t_0)^2}
\end{eqnarray}
with $t_0{=}0.71$ GeV$^2$, which is derived under the assumptions
that the Dirac form factor $F_D$ of the neutron and the isoscalar 
Pauli form factor vanish and that a dipole form is satisfactory 
for $G_M \approx \mu G_E(t)$, cf. \cite{Donnachie}. 
There are slight deviations from the dipole form in the
region $-t{<}$0.5 GeV$^2$ we are concerned with here, 
cf. for example Ref.~\cite{Belushkin}, but we neglect 
those in the present exploratory study. 
The amplitude of Eq.~(\ref{prima}) is added to the helicity 
amplitude $F_1$ of our Regge model. We show results based on 
$\Gamma(\pi^0{\to}\gamma\gamma)$=8.4 eV, i.e. the value that 
is given by the PDG as the average $\pi^0$-meson lifetime. 
The calculations are not folded with the pertinent angular 
resolution function~\cite{Braunschweig2}, because this
quantity is is not available to us for the particular 
experiment in question. 

Fig.~\ref{prima1} illustrates impressively the consequences of the Primakoff
effect. As demonstrated in the preceeding sections, 
the $\gamma p{\to}\pi^0p$ reaction amplitude can be well fixed by 
the huge set of data available at larger angles
$\theta{>}10^o$ and at different photon energies. 
This ensures that the hadronic contribution to the 
photoproduction amplitude is known quite precisely when extrapolating 
to forward angles. 
Apparently, the situation is different for measurements 
on nuclear targets. In that case the reactions at large angles are entirely
dominated by incoherent photoproduction and it is very difficult to fix 
the coherent nuclear amplitude, which contributes at forward
angles~\cite{Bauer,Trefil1,Trefil2,Sibirtsev5,Sibirtsev6}. 
Thus, $\pi^0$-meson photoproduction on the proton could offer a
promising alternative for the determination of the $\pi^0$ radiative 
decay width.
We should emphasize, however, that for the reaction on the
proton the relative phase between the hadronic part and the Primakoff
amplitude is also an unknown quantity. In the present exploratory calculation
we have simply added the latter as given in Eq.~(\ref{prima}) to our
Regge amplitude. But in a concrete application to experimental data 
one needs to determine this phase together with the $\pi^0$ radiative
decay width by a fit to differential cross sections at forward angles. 
 
\begin{figure}[b]
\vspace*{-4mm}
\centerline{\hspace*{3mm}\psfig{file=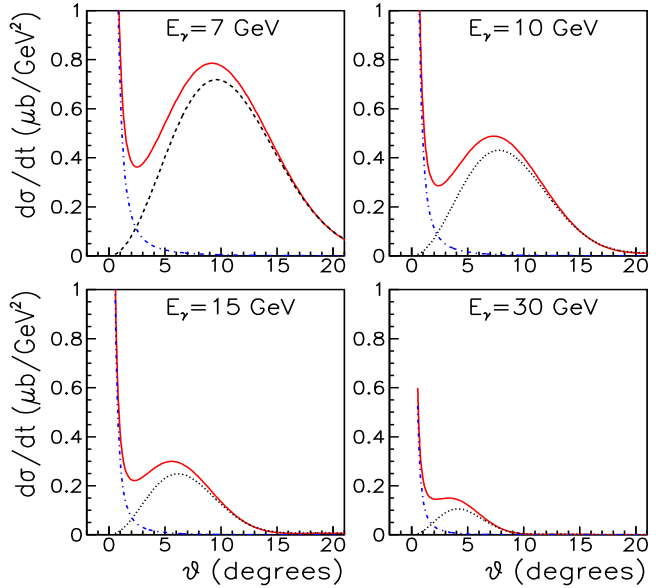,width=10.cm,height=9cm}}
\vspace*{-5mm}
\caption{Differential cross section for $\gamma{p}{\to}\pi^0{p}$ as a function
of the angle $\theta$ in the cm system shown for different photon energies
$E_\gamma$.  Same description of curves as in Fig.~\ref{prima1}.
}
\label{prim2}
\end{figure}

\begin{figure}[t]
\vspace*{-3mm}
\centerline{\hspace*{3mm}\psfig{file=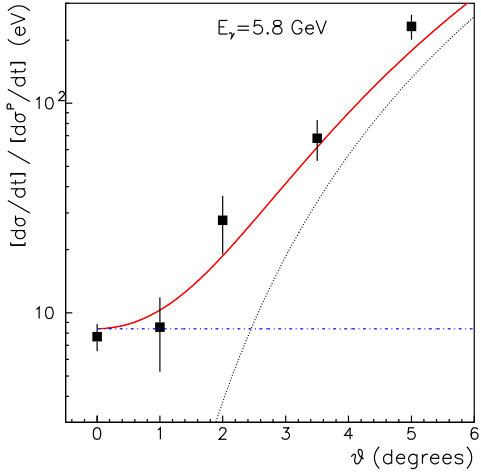,width=7.2cm}}
\vspace*{-4mm}
\caption{Differential cross section for $\gamma{p}{\to}\pi^0{p}$ 
at $E_\gamma$=5.8 GeV divided by $d\sigma^P/dt$ 
(Eq.~(\ref{obsP})).
The data are from Refs.~\cite{Braunschweig2}.
Same description of curves as in Fig.~\ref{prima1}.
}
\label{prima1a}
\end{figure}

For completeness we also show the $\gamma{p}{\to}\pi^0{p}$ differential 
cross section at forward angles for the photon energy $E_\gamma$=2~GeV, 
cf. Fig.~\ref{prima2}.  
Here the squares are data from Ref.~\cite{Braunschweig2}, triangles are the 
results from the CLAS experiment~\cite{Dugger} and circles are data from 
the CB-ELSA Collaboration~\cite{Bartholomy}. Unfortunately, the
latter recent measurements~\cite{Dugger,Bartholomy} do not cover the region 
of very forward angles.
The dotted line in Fig.~\ref{prima2} is the result of the Regge model alone
while the solid line was obtained with inclusion of the Primakoff amplitude. 
Our model reproduces the data at forward angles surprisingly well, 
but it underestimates the experimental results at $\theta{>}30^o$. 
The dashed line indicates results based on
the current solution of the GWU PWA~\cite{Arndt08}, 
which describes the data at $\theta{\ge}40^o$ but
does not reproduce the $\gamma{p}{\to}\pi^0{p}$ differential 
cross section at forward angles.

Fig.~\ref{prima2} illustrates an interesting feature. 
The excellent agreement of our Regge calculation with the available data at
forward angles could be an indication that even at such low energies the 
forward photoproduction is still dominated by $t$-channel contributions. 
If so then there would be indeed very good conditions for determining the
$\pi^0$-meson lifetime from measurements of neutral pion photoproduction 
at photon energies around $E_\gamma{\simeq}2$ GeV that are accessible 
presently at Jlab and ELSA.

Predictions for higher energies are presented in Fig.~\ref{prim2}. 
Experiments in this energy region will become feasibly once 
JLab's 12 GeV Upgrade Project will be completed.

Finally, let us provide another view on the present situation.  
In Fig.~\ref{prima1a} we show again the available cross-section data at 
5.8~GeV \cite{Braunschweig2}, but divide the data and the curves by the
contribution of the pure Primakoff amplitude with a normalization
so that the result at zero angle coincides with the $\pi^0$ 
decay width, i.e. we divide by 
\begin{eqnarray}
\frac{d\sigma^P}{dt}&=&\frac{1}{32\pi}\left[ \frac{
t|\hat F^P|^2}{(t-4m^2_p)}\right] \ . 
\label{obsP}
\end{eqnarray}
Note that a logarithmic scale is used for the
ordinate. On this plot one can see down to which angles the 
results are still dominated by the hadronic amplitude. It 
is obvious that an appropriate set of data points below 3 degrees, 
say, and with high precision would allow an extrapolation to 
zero degrees and, thus, a determination of $\Gamma(\pi^0{\to}\gamma\gamma)$.
The presently available data are too sparse and too inaccurate for
performing such an extrapolation reliably. 

\section{Summary}

In the present paper we performed a global analysis of the world 
data on the reactions $\gamma{p}{\to}\pi^0p$ and $\gamma{n}{\to}\pi^0n$ 
for photon energies from 3 to 18~GeV within the Regge approach. 
In this region resonance contributions are expected to be negligible so 
that the available experimental information on differential cross sections 
and single- and double polarization observables at $-t{\leq}2$ GeV$^2$
allows us to determine the reaction amplitude reliably.
The Regge model was constructed by taking into account both pole and cut 
exchange $t$-channel helicity amplitudes and includes the $\rho$, 
$\omega$ and $b_1$ trajectories. The model parameters such as the
helicity couplings were fixed by a fit to the available data in the
considered $E_\gamma$ and $t$ range. 

An excellent overall description of the available data was achieved, 
indicating that for the energy and $t$ range in question single 
pion photoproduction is indeed dominated by nonresonant 
contributions. 
The model amplitude was then used to predict observables for photon 
energies below $3$ GeV. A detailed comparison with recent data from
the CLAS (JLab) and CB-ELSA (Bonn) Collaborations in that energy
region was presented. It turned out that the resulting differential 
cross sections for $\gamma{p}{\to}\pi^0p$ were still in reasonable
agreement with those new data down to $E_\gamma \approx$ 2.45 GeV
for $-t{\leq}2$ GeV$^2$, while the very forward data were reproduced
even down to photon energies as low as 2~GeV.

Since our Regge amplitude works so well for forward angles, even
at very low energies, 
we utilized it to explore the prospects for determining the $\pi^0$ 
radiative decay width via the Primakoff effect from the reaction 
$\gamma{p}{\to}\pi^0p$. 
Those calculations indicate that corresponding measurements on a
proton target could be indeed promising. But, evidently, the 
precision to which the decay width can be determined will 
depend crucially on the number of data points that one can collect 
at very small angles and on the accuracy and the angular resolution 
one can achieve. 

\subsection*{Acknowledgements}
We acknowledge fruitful discussions with A. Bernstein, W. Chen, M. Dugger, 
L. Gan, H. Gao, A. Gasparian, J. Goity, I.~Strakovsky, and U.~Thoma.
This work is partially supported by the Helmholtz Association through funds
provided to the virtual institute ``Spin and strong QCD'' (VH-VI-231), by \,
the European Community-Research
Infrastructure Integrating Activity ``Study of Strongly Interacting
Matter'' (acronym HadronPhysics2, Grant Agreement no. 227431) under the
Seventh Framework Programme of EU,
and by DFG (SFB/TR 16, ``Subnuclear
Structure of Matter''). 
This work was also supported in part by U.S. DOE
Contract No. DE-AC05-06OR23177, under which Jefferson Science Associates, LLC,
operates Jefferson Lab. A.S. acknowledges support by the JLab grant
SURA-06-C0452 and the COSY FFE grant No. 41760632 (COSY-085). 

\newpage
\section{Appendix A}
Utilizing the relations of Ref.~\cite{Barker} the 
$\gamma{N}{\to}\pi{N}$ observables analysed in our study are given
in terms of the amplitudes $F_i$ ($i{=}1,...,4)$ by~\cite{Barker1}
\begin{eqnarray}
\frac{d\sigma}{dt}\phantom{\Sigma}&=&\frac{1}{32\pi}\left[ \frac{
t|F_1|^2-|F_3|^2}{(t-4m^2_N)} +|F_2|^2-t|F_4|^2\right], 
\label{obs1} \\
\frac{d\sigma}{dt}\Sigma&=&\frac{1}{32\pi}\left[ \frac{
t|F_1|^2-|F_3|^2}{(t-4m^2_N)} -|F_2|^2+t|F_4|^2\right],
\label{obs2} \\
\frac{d\sigma}{dt}T&=&\frac{\sqrt{-t}}{16\pi}\,\, {\rm Im} \left[ \frac{
-F_1 F_3^\ast}{(t-4m^2_N)} + F_4F_2^\ast \right],
\label{obs3} \\
\frac{d\sigma}{dt}P&=&\frac{\sqrt{-t}}{16\pi}\,\, {\rm Im} \left[ \frac{
-F_1 F_3^\ast}{(t-4m^2_N)} - F_4F_2^\ast \right],  \, 
\label{obs4} \\
\frac{d\sigma}{dt}G&=&\frac{{\rm Im} \left[ 
t F_4(F_3^\ast {-} 2m_NF_1^\ast) {+}F_2(tF_1^\ast {-} 2m_NF_3^\ast) \right]} 
{16\pi (t{-}4m^2_N)}, \, 
\label{obs6}\\
\nonumber 
\frac{d\sigma}{dt}H&=&\frac{\sqrt{{-}t}\, {\rm Im} \left[ 
F_2(F_3^\ast {-} 2m_NF_1^\ast) {+}F_4(tF_1^\ast {-} 2m_NF_3^\ast) \right]} 
{16\pi (t{-}4m^2_N)}. \,\,\,\,\,
\\
\label{obs8}
\end{eqnarray}
In order to account for the correct behavior at very small angles
$t$ should be replaced by $t-t_{min}$ in the above formulae, where 
$t_{min} = - (m_{\pi^0} / 2 E_\gamma)^2$.

\section{Appendix B}
Here we provide the relation between the $t$-channel helicity
amplitudes $F_i$ 
and the $s$-channel helicity amplitudes $S_1$, $S_2$, $N$ and $D$.
Following Wiik's abbreviations~\cite{Wiik},  $S_1$ and $S_2$ are single 
spin-flip amplitudes, 
$N$ is the spin non-flip and $D$ is the double spin-flip 
amplitude, respectively.
The asymptotic crossing relation, which is useful for the analytical
evaluation of the helicity amplitudes, is given by
\begin{eqnarray}
\nonumber 
\!\!\!\left[
\begin{matrix}{F_1 \crcr F_2 \crcr F_3 \crcr F_4}
\end{matrix}
\!\right]\!\!{=}\frac{-4\sqrt{\pi}}{\sqrt{-t}}\!\!
\left[\begin{matrix}{\!2m_N & \sqrt{-t} & -\sqrt{-t} & \!\!2m_N \crcr 
0 & \sqrt{-t} & \sqrt{-t} & 0 \crcr 
t & \!\!2m_N\sqrt{-t} & \!\!{-}2m_N\sqrt{-t} & t \crcr 
1 & 0 & 0 & {-}1 }
\end{matrix}\!\!
\right]\!\!
\left[
\begin{matrix}{S_1 \crcr N \crcr D \crcr S_2}
\end{matrix}
\right] . \\
\label{matr1}
\end{eqnarray}

Note that Eq.~(\ref{matr1}) is appropriate only at $s{\gg}t$, since it does not 
account for higher order corrections that are proportional to $t/4m^2_N$. 
The amplitudes $F_i$ are related to the usual CGLN 
invariant amplitudes $A_i$ \cite{Chew} by 
\begin{eqnarray}
F_1&=&-A_1+2m_NA_4~, \nonumber \\
F_2&=&A_1+tA_2~, \nonumber \\
F_3&=&2m_NA_1-tA_4~, \nonumber \\
F_4&=&A_3 \ . 
\label{invari}
\end{eqnarray}
Expressions for the experimental observables in terms of the
amplitudes $A_i$ are listed, for instance, in Ref.~\cite{Berends}. 
The often used multipole amplitudes can be constructed from
the helicity amplitudes using the relations given in
Refs.~\cite{Ball,Kramer1,Said1}.

\end{document}